\RequirePackage{fix-cm}
\documentclass[smallcondensed]{svjour3}     
\smartqed  
\usepackage{graphicx}
\usepackage{natbib}
\journalname{Celestial Mechanics and Dynamical Astronomy}
\begin{document}

\title{Three dimensional structure of mean motion resonances beyond Neptune
	\thanks{to be published in CMDA topical collection on Trans-Neptunian Objects }
}


\author{Tabar\'{e} Gallardo}


\institute{T. Gallardo \at
             Facultad de Ciencias, Universidad de la Rep\'{u}blica, Igu\'{a} 4225, 11400 Montevideo, Uruguay \\
              \email{gallardo@fisica.edu.uy}    
}

\date{Received: July 22, 2019/ Accepted: December 10, 2019}

\maketitle

\begin{abstract}
	We propose a semianalytical method for the calculation of widths, libration centers and small amplitude libration periods  of the mean motion resonances
	$k_p$:$k$ in the framework of the circular restricted three body problem valid for arbitrary eccentricities and inclinations. 
	Applying the model to the trans Neptunian region (TNR) we obtain several atlas of resonances between 30 and 100 au showing their domain in the plane $(a,e)$ for different orbital  inclinations.
	The resonance width may change substantially when varying the argument of the perihelion of the resonant object and in order to take into account these variations we introduce the concept of resonance fragility.
	Resonances 1:$k$ and 2:$k$ are the widest, strongest, most isolated ones and with  lower fragility for all interval of inclinations and eccentricities. 
	We discuss about the existence of high $k_p$:$k$ resonances.	
	We analyze the distribution of the resonant populations inside  resonances
	1:1, 2:3, 3:5, 4:7, 1:2 and 2:5.
We found that the populations are in general located near the regions 
 of the space $(e,i)$ where the resonances are wider  and less fragile with the notable exception of the population inside the resonance 4:7 and in a lesser extent the population inside 3:5 which are shifted to lower eccentricities.

\end{abstract}

\section{Introduction}
\label{intro}

The resonant structure of the TNR was explored by numerical integrations of test particles \citep[for example]{1993ApJ...406L..35L,1995AJ....110.3073D,Malhotra1995a,2000Icar..147..205M}
and by application of analytical theories that depend on the adopted approach for the resonant disturbing function. Analytical expansions of the  disturbing function \citep[for example]{Beauge1996,2000Icar..147..129E} are limited to some interval of eccentricities and/or inclinations. Analytical expansions around arbitrary specific points of the  phase space (also called \textit{asymmetric expansions}) allowed the study of the resonant motion around the center of the expansion \citep{1989A&A...225..541F,1995CeMDA..62..145G,1998A&A...329..339R} and they were applied  to understand the dynamics of the asteroidal resonances mostly.
 Semianalytical methods, that means analytical theories based on the numerical evaluation of the disturbing function, allowed a very precise description of the planar mean motion resonances (MMRs) in the asteroid belt and in the TNR  \citep[and subsequent references]{1964SAOSR.149.....S,Moons1993a,Morbidelli1995}.
Resonance's properties for planar direct orbits were very well described since then. 
Retrograde resonances appeared in the eighties related to studies on dynamical evolution of comets \citep{1985PAZh...11..924E,1986ESASP.250b.413C}, but 
probably because it sounded very unlikely,
only in the last years the planar retrograde 
\citep{2012MNRAS.424...52M,2013CeMDA.117..405M}
and then the general inclined resonance problem was started to be studied systematically \citep{2013MNRAS.436L..30M,2015MNRAS.446.1998N,2016CeMDA.125...91M,2017MNRAS.467.2673N,2017MNRAS.472L...1M,2018CeMDA.130...29V,10.1093/mnras/stz1422}. Some of these studies  provide a general picture of the resonances but limited to some interval of eccentricity and/or inclination.

An analytical expression for the  resonant disturbing function for arbitrary $(e,i)$ has in general several  terms that must be taken into account globally in order to have a complete picture of the resonance.
\cite{2017MNRAS.471.2097N} and \cite{2018MNRAS.474..157N} presented a  theory that allows to find specific terms of the analytical expansion for arbitrary spatial resonances.
The theory by \cite{10.1093/mnras/stz1422} is very similar to the one presented by \cite{2018MNRAS.474..157N} but arranging the resonant terms in a more compact way providing a more global description of the whole resonance and not limited to individual resonant terms. Both theories are valid in all interval of inclinations but limited to  $e\leq 0.5$ due to convergence problems in the series expansions. In a different approach
and by means of  a numerical evaluation of the resonant disturbing function  \cite{2006Icar..184...29G,2019Icar..317..121G}
 calculated the resonance's strength and provided a general picture of all kind of resonances in therms of strengths. 
Both approaches by \cite{10.1093/mnras/stz1422} and \cite{2006Icar..184...29G,2019Icar..317..121G} put in evidence that the particle's argument of the perihelion is crucial for the definition of the strength and width of a resonance in the spatial case.

We present here a semianalytical model based on the numerical evaluation of the resonant disturbing function, which assumes some approximations that simplify greatly the theory providing a very fast method for automatically computing 
equilibrium points, libration periods and widths
 of  MMRs for orbits with arbitrary eccentricities and inclinations. 
Moreover, we introduce a new concept: the fragility of resonances.
In section \ref{method} we explain and test the model.
In section \ref{proper}  we extensively apply the model  to the TNR obtaining  several atlas containing hundreds of resonances and we discuss about the existence of 
high $k_p$:$k$ resonances.
In section \ref{respop} we analyze six known populations of resonant trans Neptunian objects (TNOs).
We end with a summary in section \ref{discu}.

\section{A model for spatial MMRs}
\label{method}

\subsection{Resonant Hamiltonian}

Let us consider the MMR that we note as
 $k_p$:$k$,
 which corresponds to the commensurability
 $k_pn_p\simeq kn$, being $k_p,k$ positive integers 
 that do not have common divisors
  and $n_p,n$ the mean motions 
 of the planet and the TNO respectively.
The corresponding critical angle is given by
 \begin{equation}\label{crit}
 \sigma =  k\lambda -k_p\lambda_p   + (k_p - k)\varpi
 \end{equation}
 where subindex $p$ denotes  planet.
Note that using this notation 
 3:1, for example, is an interior resonance and 1:3 an exterior resonance
 and it is not necessary to specify whether it is interior or exterior or whether it is direct or retrograde.
 The last term for  $\sigma$ in Eq. (\ref{crit}) can be a combination of $\varpi$ and $\Omega$ (longitudes of the perihelion and ascending node respectively) but that is not relevant for our purposes as we will explain. 
Following, for example, \cite{2002aste.conf..379N} or \cite{2016CeMDA.126..369S} the semi secular Hamiltonian obtained
eliminating the short period terms depending on 
 $\lambda$ or $\lambda_p$, but not on $\sigma$, 
is
\begin{equation}\label{fham}
\mathcal{H}(a,e,i,\omega,\sigma) = -\frac{\mu}{2a} -n_p\frac{k_p}{k}\sqrt{\mu a}  - \mathcal{R}(a,e,i,\omega,\sigma)
\end{equation}
where $\mu = GM_{\odot}$. 
Note that the dependence with
$e,i,\omega$ is through the resonant disturbing function $\mathcal{R}$.
As $\mathcal{H}$ 
does not depend explicitly with the time it is conserved and
 the solutions occur in surfaces defined by $\mathcal{H}(a,e,i,\omega,\sigma) = constant$.
Several analytical developments of  $\mathcal{R}(a,e,i,\omega,\sigma)$ 
have been proposed, each one valid in some interval of the orbital elements.
We will adopt here the approximation given by \cite{2006Icar..184...29G,2019Icar..317..121G} where, for the resonance $k_p$:$k$,   $\mathcal{R}$ is numerically evaluated assuming fixed values for $(a\equiv a_0,e,i,\omega)$: 
\begin{equation}\label{rmean}
\mathcal{R}(\sigma)=\frac{1}{2\pi k}\int_{0}^{2\pi k}R(\lambda_p,\lambda(\lambda_p,\sigma))d\lambda_p
\end{equation}
where $a_0$ is the nominal value for the exact resonance and $R$ is the disturbing function of the planet with mass $m_p$ and heliocentric position $\mathbf{r}_p$ on the particle with heliocentric position $\mathbf{r}$:
\begin{equation}\label{defr}
R=G m_p\Bigl(\frac{1}{\mid \mathbf{r}_p-\mathbf{r}\mid} - \frac{\mathbf{r}\cdot\mathbf{r}_p}{r^3_p}
\Bigr)
\end{equation}
The mean longitude of the asteroid, $\lambda$,  in  (\ref{rmean}) is expressed as a function of $(\lambda_p,\sigma)$ according to Eq. (\ref{crit}).
This numerical averaging is the same proposed by \cite{1964SAOSR.149.....S} but assuming  fixed $(a,e,i,\omega)$ in the calculation of the integral  (\ref{rmean}).
The assumption of fixed $\omega$ does not introduce any spurious result  because  $\omega$ varies in very long time scales. On the other hand, the elements $(a,e,i)$ do vary a little during one resonant libration but, as we will show, their variations do not introduce relevant changes in the numerical calculation of the equilibrium points, libration periods and resonance widths.
Then, the approximation  assumed implies that the resonant Hamiltonian have two variables $(a,\sigma)$ and it depends also on the fixed parameters $(e,i,\omega)$. The solutions will be level curves $\mathcal{H}(a,\sigma)$, and analyzing them
 we can identify  the stable and unstable equilibrium points, the librations around the stable points and the separatrices as we will explain below.
The level curves of $\mathcal{H}=constant$ are calculated using
\begin{equation}\label{mapa}
\mathcal{H}(a,\sigma) = -\frac{\mu}{2a} -n_p\frac{k_p}{k}\sqrt{\mu a}  - \mathcal{R}(a_0,\sigma)
\end{equation}

\begin{figure}
	\includegraphics[width=1.\textwidth]{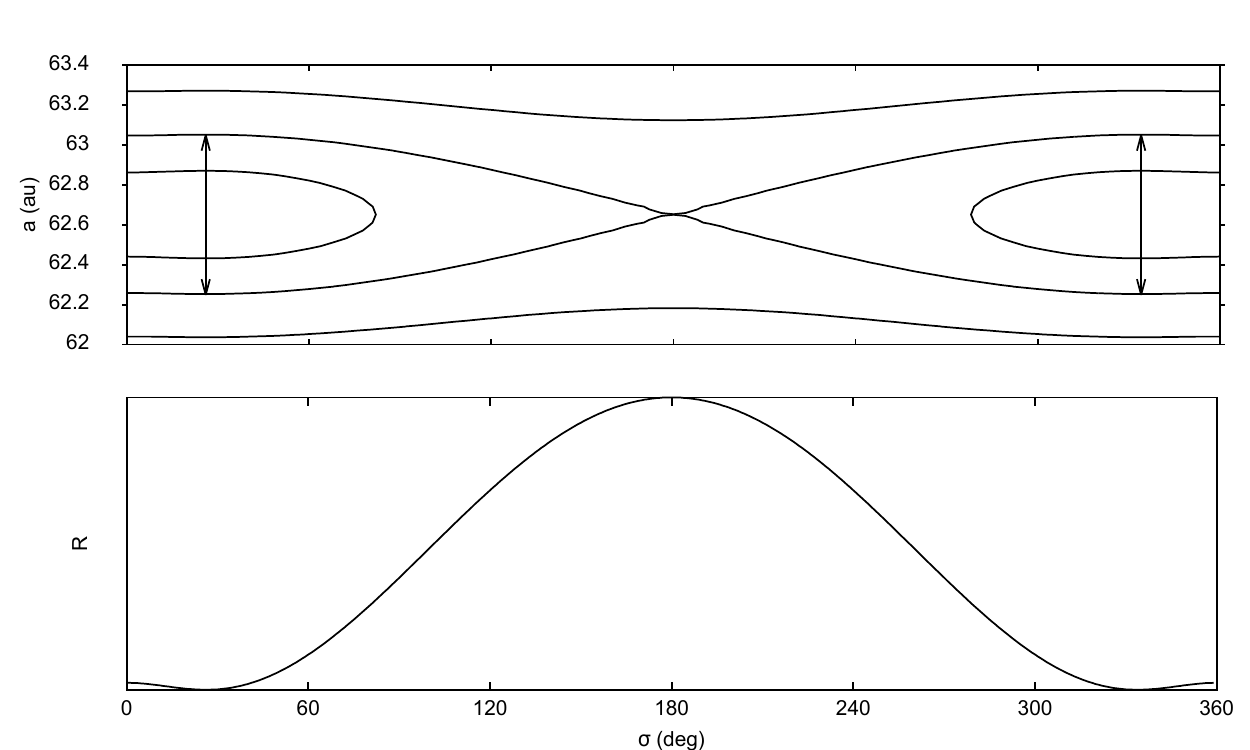}
	\caption{Resonance 1:3 with Neptune for an orbit with $e=0.7, i=90^{\circ}, \omega=0^{\circ}$. Top: some level curves for $\mathcal{H}(a,\sigma)$  given by Eq. (\ref{mapa}) including the separatrix. 
		 Bottom: corresponding $\mathcal{R}(a_0,\sigma)$ given by Eq. (\ref{rmean}). The stable equilibrium points are the minimum for $\mathcal{R}$		 
		  located at $\sigma=26^{\circ}$ and $\sigma= 334^{\circ}$.
		  The unstable equilibrium points are located at $\sigma=0^{\circ}$ and $\sigma=180^{\circ}$.		
		The separatrix passing by the unstable equilibrium point at $\sigma=180^{\circ}$ and $a_0=62.63$ au defines the total width of the resonance (see section \ref{rewi}), indicated with arrows, which turns out to be 0.8 au.}
	\label{levelsi90}   
\end{figure}

\subsection{Equilibrium points and libration periods}
\label{eqpo}

To obtain the equations for $(a,\sigma)$
we should first change to the canonical variables  $(\Sigma, \sigma)$ where $\Sigma=\sqrt{\mu a}/k$.
The canonical equations are
\begin{equation}\label{bigsig}
\frac{d\Sigma}{dt} = -\frac{\partial \mathcal{H}}{\partial \sigma}
\end{equation}
\begin{equation}\label{sigdot}
\frac{d\sigma}{dt} = \frac{\partial \mathcal{H}}{\partial \Sigma}
\end{equation}
From Eq. (\ref{bigsig}) operating we obtain
\begin{equation}\label{adot}
\frac{da}{dt} = \frac{2k}{\sqrt{\mu /a}}\frac{\partial\mathcal{R}}{\partial\sigma}
\end{equation}
from which we conclude that the dependence of $\mathcal{R}$ with $\sigma$ defines the dynamical behavior of the resonance because 
the stronger the dependence  of $\mathcal{R}$ with $\sigma$ the
larger the rate of change of $a$ due to the resonant motion.
The equilibrium points are defined by the condition
\begin{equation}\label{equi}
\frac{d\sigma}{dt} = \frac{d a}{dt}=0
\end{equation}
Using this condition, 
from Eq. (\ref{crit}) assuming $d\varpi/dt=0$ it follows that the equilibrium points are at
$a=a_0$. But using Eqs. (\ref{equi})
and  (\ref{adot}) we have that the equilibrium points verify
\begin{equation}
 \frac{\partial\mathcal{R}}{\partial\sigma} =0
\end{equation}
In Fig.  \ref{levelsi90} we show an example where the equilibrium points are located at  $\sigma=$ 0$^{\circ}$, 26$^{\circ}$, 180$^{\circ}$ and 334$^{\circ}$.
Being 
 $(\Sigma_0,\sigma_0)$ 
 an equilibrium point in canonical variables, if we consider some small displacement $(S,s)$, using the canonical equations we can obtain the first order expansions 
 \begin{equation}
 \frac{dS}{dt} =  - \mathcal{H}_{\sigma\sigma}s - \mathcal{H}_{\sigma\Sigma}S
 \end{equation}
 \begin{equation}
 \frac{ds}{dt} =  \mathcal{H}_{\Sigma\sigma}s + \mathcal{H}_{\Sigma\Sigma}S
 \end{equation}
where subscripts in  $\mathcal{H}$ mean partial derivatives. Looking for solutions of the type $S=A\exp(2\pi t/T)$ and $s=B\exp(2\pi t/T)$
	it is straightforward to prove that oscillations only occur with a
 libration period, $T$, in years given by
\begin{equation}
T=\frac{a}{k}\frac{2\pi}{\sqrt{3\mathcal{R}_{\sigma\sigma}}}
\end{equation}
where $\mathcal{R}_{\sigma\sigma}$ is the second derivative calculated numerically at the stable equilibrium point.

\subsection{Resonance widths}
\label{rewi}

The resonance's half width
 $\Delta a$ 
 is equal to the difference between $a_0$ and $a_{sep}$ where $a_{sep}$ is defined by the separatrix such that
 \begin{equation}\label{hh}
 \mathcal{H}(a_{sep},\sigma_s)=\mathcal{H}(a_{0},\sigma_u)
 \end{equation}
 being $\sigma_s$ and $\sigma_u$ the stable and unstable equilibrium points.
The total width is twice  $\Delta a$ and in Fig.  \ref{levelsi90} is shown with vertical arrows. 
 Be $\Delta \mathcal{H}=\mathcal{H}(a_{sep},\sigma_s)-\mathcal{H}(a_{0},\sigma_s)$, then  we can approximate
\begin{equation}
\Delta \mathcal{H}=\frac{\partial \mathcal{H}}{\partial a} \Delta a + 
\frac{\partial^2 \mathcal{H}}{\partial a^2} \frac{(\Delta a)^2}{2} + \dots
\end{equation}
Evaluating the derivatives at the stable equilibrium point and using (\ref{hh})
we have
\begin{equation}
\Delta \mathcal{H}=\mathcal{H}(a_{0},\sigma_u)-\mathcal{H}(a_{0},\sigma_s)\simeq
\frac{\partial^2 \mathcal{H}}{\partial a^2} \frac{(\Delta a)^2}{2} 
\end{equation}
The left hand is
\begin{equation}
 \mathcal{R}(\sigma_s)-\mathcal{R}(\sigma_u)=-\Delta \mathcal{R}
\end{equation}while
$$\frac{\partial^2 \mathcal{H}}{\partial a^2}=-\frac{3}{4}n^2 $$
then the half width of the resonance expressed in au is
\begin{equation}
\Delta a \simeq \frac{\sqrt{8/3}}{n}\sqrt{\Delta \mathcal{R}}
\end{equation}
where $\Delta \mathcal{R}$ is the maximum amplitude of $\mathcal{R}(\sigma)$.
Then for a specific resonance with a defined planet the method consists in, given $(e,i,\omega)$, calculate numerically the function $\mathcal{R}(\sigma)$ and deducing numerically $\Delta \mathcal{R}$ and $\mathcal{R}_{\sigma\sigma}$ at the stable equilibrium points  in order to calculate $\Delta a$ and the periods of the small amplitude librations. 
As we assume $\varpi$ and $\Omega$ constants it is irrelevant which
combination of these angles we use in (\ref{crit})
 because all them will generate the same  $\Delta \mathcal{R}$ and  $\mathcal{R}_{\sigma\sigma}$.
Up to now this method is analogue to the analytical method given by \cite{10.1093/mnras/stz1422} but here we calculate  $\mathcal{R}$ numerically while in  \cite{10.1093/mnras/stz1422}   an analytical expansion is used and consequently it is not applicable for very high eccentricities. Both methods calculate widths using $\mathcal{R}$ evaluated at the equilibrium points and are unable to distinguish asymmetries between the left and right limits of the resonances. Those asymmetries are especially noticeable in some first order resonances with Jupiter \citep{2002aste.conf..379N}.

\subsection{Calculation of $\Delta \mathcal{R}$ in close encounters}

The calculation of the resonance width is a fundamental problem in resonance dynamics. Usually, the width is defined by the separatrix in an analogue procedure as we have done here. But the region close to the separatrix is chaotic 
and chaos  is especially important when resonances are so wide that they overlap. In these cases the widths deduced from separatrices 
are larger than the regions where the oscillations are really stable.
	Also, for sufficiently eccentric  orbits having large amplitude librations, a close encounter between the object and the planet can occur, generating a peak in $\mathcal{R}$ and disrupting the resonant motion. 
To avoid these flaws and in order to have reliable widths for the planar case
 \cite{Malhotra1995a}, \cite{2018AJ....156...55M} and \cite{Lan2019} defined the widths only for stable librations obtained by means of numerical explorations using Poincaré surface of sections.
For the same reason \cite{2006Icar..184...29G} defined the  strength of the resonance as $SR=<\mathcal{R}>-\mathcal{R}_{min}$, a parameter that is not strongly affected by the maximum peaks of $\mathcal{R}$ generated in situations of close encounters.

Using our approach, in the process of calculation of $\mathcal{R}(\sigma)$ for a given $\sigma$,
when varying $\lambda$ and $\lambda_p$,
the particle may be placed in a configuration of close encounter with the planet. If there is no collision it is possible to calculate $\mathcal{R}(\sigma)$ with enough precision, but it will be an unstable configuration and in the real world the resonant motion will be broken. Then, following the criteria of stable librations, in order to calculate $\Delta \mathcal{R}$ to obtain reliable maximum resonance widths, we do not take into account values of $\mathcal{R}$ obtained in circumstances of close encounters. After some numerical experiments comparing our predicted widths with the results of numerical integrations we found that a safe distance is $3R_{H}$, where $R_H$ is the planetary Hill's radius. 
A resonant TNO encountering Neptune at less than $3R_{H}$ brakes the resonant configuration. We have also found that the limit has some dependence with the orbital inclination. For near zero inclination orbits the actual limit probably is between 3 and 4$R_{H}$ but for high inclinations is  closer to 2$R_{H}$. The disruption of the resonant motion depends on the minimum distance to the planet but also on the relative velocity which strongly depends on the orbital inclination. So,  $3R_{H}$ is a compromise in order to have an idea of the safe maximum resonance widths in situations of close encounters.
We remark that, theoretically, libration amplitudes can be larger but most probably unstable. 
In our model we do not consider the superposition of resonances as a limitation for the resonance width. If that situation exists, it will be evident when calculating a series of neighboring resonances.
 We have written a code in FORTRAN for
computing  $\mathcal{R}(\sigma), \Delta \mathcal{R}$, the total width of the resonance, the location of the equilibrium points and corresponding libration periods of the small amplitude oscillations. It is not difficult to generalize the algorithm to the case of an eccentric planet. It can be downloaded from
www.fisica.edu.uy/$\sim$gallardo/atlas/ra.
In the next subsection we present some tests of our model in extreme situations.

\subsection{Testing the model with dynamical maps}
\label{sec:2}

\begin{figure}
  \includegraphics[width=1\textwidth]{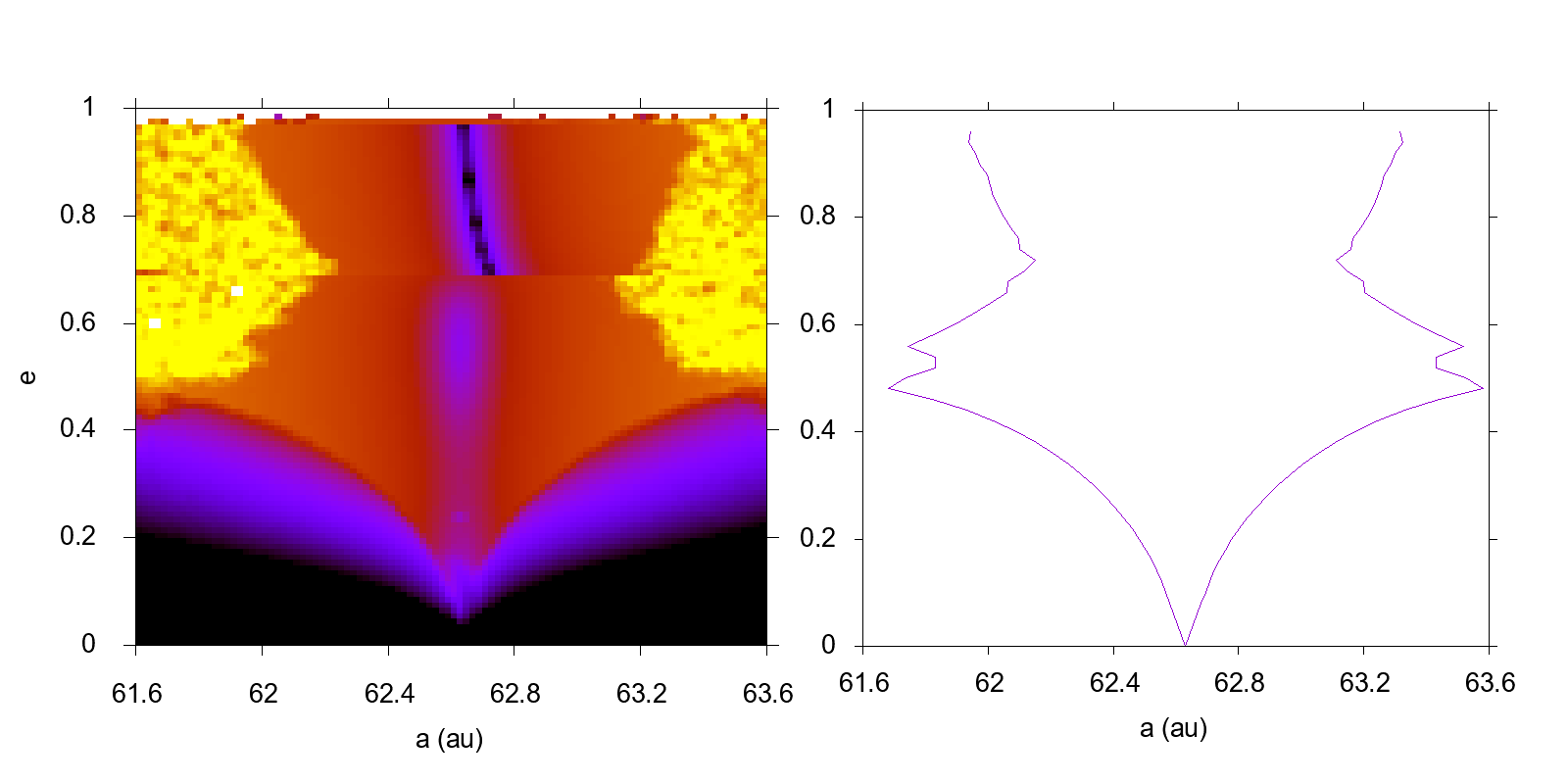}
\caption{Resonance 1:3 with Neptune. Left: dynamical map showing in color scale the logarithm of $\Delta a$  in au as function of the initial $(a,e)$ for test particles with initial $i=0^{\circ}$. Yellow regions correspond to $\Delta a >10$ au and blue and black to  $\Delta a < 0.1$ au. For $0<e<0.68$ the initial critical angle was taken $\sigma=300^{\circ}$ and for 	 $e>0.68$ was taken $\sigma=0^{\circ}$. For $e>0.51$ there are intersections with  Neptune's orbit. 
The red region, corresponding to  $\Delta a$ of the order of some au, defines the limits of the stable domain of the resonance.
	Right:
 the limits predicted with the model.}
\label{map1to3i0}  
\end{figure}

\begin{figure}
	\includegraphics[width=1\textwidth]{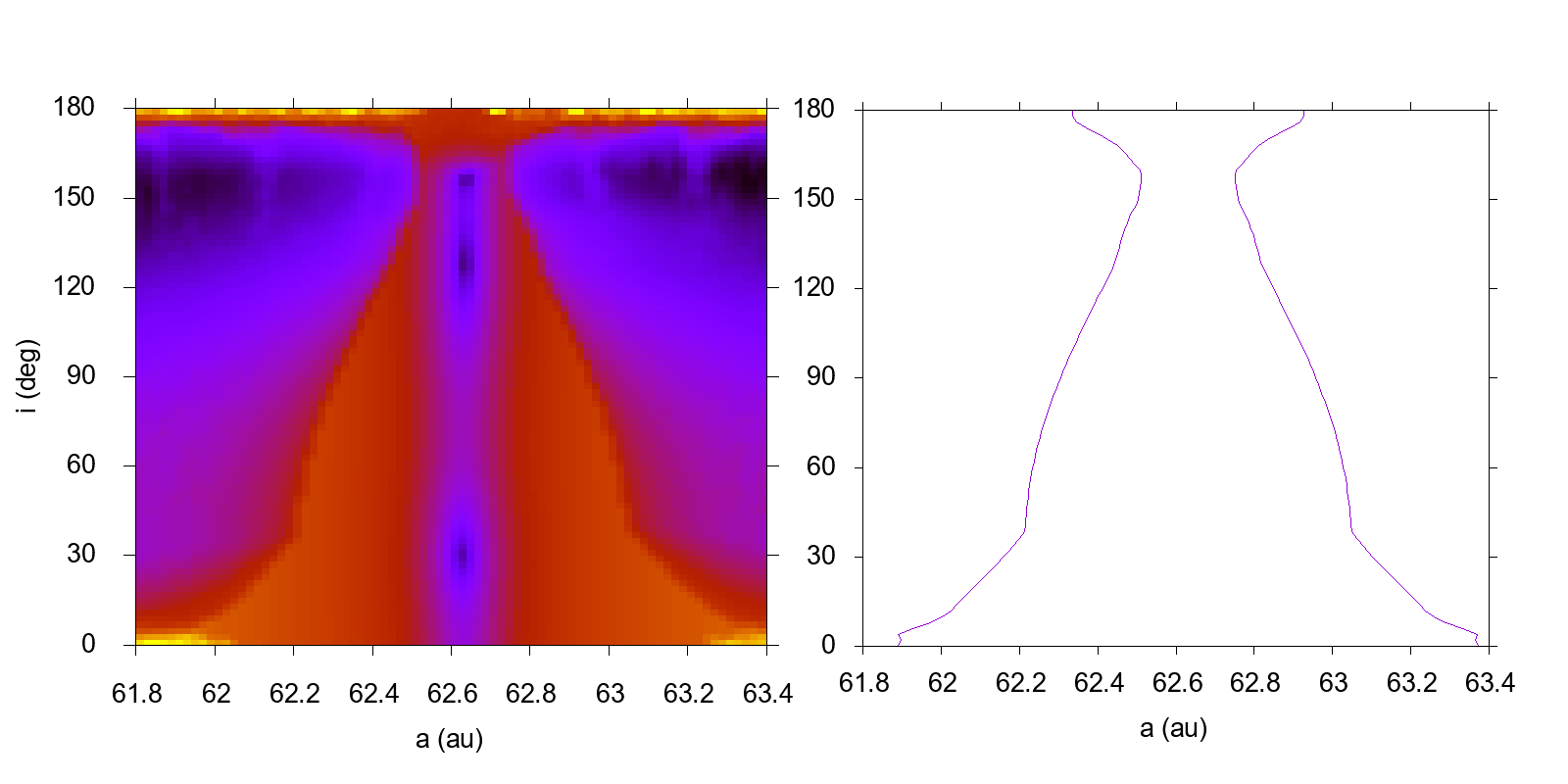}
\caption{Resonance 1:3 with Neptune. Left: dynamical map showing in color scale the logarithm of $\Delta a$  in au as function of the initial $(a,i)$ for test particles with itinial $e=0.6$ and $\omega=0^{\circ}$. Yellow regions correspond to $\Delta a >10$ au and blue and black to  $\Delta a < 0.1$ au. The initial critical angle was taken  $\sigma=300^{\circ}$. The red region, corresponding to $\Delta a$ of the order of some au, defines the limits of the stable domain of the resonance.
Right: the limits predicted with the model.}
	\label{map1to3e0p6}     
\end{figure}

We have tested our model comparing the predicted widths with the ones that can be deduced by means of dynamical maps calculated with the numerical integrations of the exact equations of motion considering the Sun, Neptune in circular and zero inclination orbit and massless particles in arbitrary orbits. We have found a very good agreement in very different circumstances and we will illustrate with two extreme cases related to the resonance 1:3 involving collision with Neptune.
The first case is the study of the resonance widths in the space $(a,e)$ for orbits with $i=\omega=0^{\circ}$. These conditions imply that for $e>0.51$ there will be a collision for some value of  $\sigma$. We calculated dynamical maps taking a grid of initial conditions covering $61.6<a<63.6$ au and $0<e<0.98$ and we computed the mean baricentric semimajor axes after 10 orbital periods and then the  variation  $\Delta a$ of the mean values after 200 orbital periods. Using this methodology we eliminate short period oscillations of $a$ and we can distinguish the orbital changes due to the librations.
The resulting dynamical map strongly depends on
 the chosen initial value of $\sigma$ in the numerical integrations. For different initial  $\sigma$ we will obtain different libration amplitudes so, for different intervals in $e$, we choose different initial  $\sigma$ according to the locations of the equilibrium points, so that the resonant regions   obtained in the map are the widest possible. The result is shown in Fig.  \ref{map1to3i0} left panel, where  black and blue regions of the map correspond to minimum changes typical of secular evolutions, red corresponds to oscillations due to the resonant motion and yellow corresponds to large changes due to disruption of the resonance. The domain of the stable resonance is the red region. In the yellow regions for $e>0.5$ the orbits abandon the resonance due to close encounters with Neptune.  The dark regions near the nominal value $a_0$ are due to small amplitude oscillations around the equilibrium points inside the resonance. 
 In Fig.  \ref{map1to3i0} right panel we show the limits computed by our algorithm calculating $\Delta \mathcal{R}$ rejecting values of $\mathcal{R}$ obtained with close encounters with distances less than  $3R_{H}$. Even for the extreme situations when $e>0.5$ our algorithm is capable to detect quite correctly the limits of the stable borders of the resonance. There is a very good match
 with Fig.  2 second panel of \cite{Lan2019} where the resonance widths were obtained by Poincaré sections.
Another example is shown in Fig.  \ref{map1to3e0p6}. 
It is the same resonance but studied in the plane  $(a,i)$ for $e=0.6$ 
and  $\omega=0^{\circ}$. The map in the left panel was calculated taking initial $\sigma=300^{\circ}$ in order to obtain the widest librations. The right panel corresponds to the calculated limits according to our model showing a very good agreement. 
 We have also found a very good agreement between the limits  for the resonance 2:5 given by our model and Fig.  3 by \cite{2018AJ....156...55M}. 
Then, the model gives good approximations to the widths of the stable resonant orbits.
Nevertheless, in order to represent more exactly the actual limits of the resonances in situations of close encounters, a fine tunning of the algorithm can be done adjusting the tolerance to the close encounters with the planet when calculating  $\Delta \mathcal{R}$. Retrograde orbits probably can tolerate close encounters up to 2$R_H$ while for direct orbits the limit could be shifted to  4$R_H$.

The weakest part of this model is assuming that $(e,i)$ are fixed during the librations, assumption that was avoided in the literature since the beginning of the application of semianalytical methods for MMRs because the aim was precisely
to find the time evolution of the eccentricity. 
For this reason this model is unable to describe the time evolution of $(e,i)$ but it is useful   for the determination of the resonance widths which by definition are the maximum $\Delta a$ of the resonance's domain in the space $(a,e,i)$ for given, fixed, $(e,i)$.

We have also applied the model to resonances with Jupiter, where the librations of $(e,i)$ are the largest in the Solar System, and comparing with the widths deduced from dynamical maps we have found an excellent agreement in absence of situations of close encounters.
	For example, we compared our calculated widths with the ones presented by \cite{10.1093/mnras/stz1422} for resonances 2:1 and 3:1 with Jupiter for $e<0.5$ obtaining a perfect agreement.
But, when the eccentricity is large enough to allow close encounters sometimes it is necessary to adjust the criteria of rejection of data from 3$R_H$ to even 1$R_H$ according to the case. 
Just for illustration, the total width of the resonance 3:2 with Jupiter for $i=0^{\circ}$ and $e=0.5$ is 0.33 au according to the distance between its separatrices (see for example \cite{2002aste.conf..379N}). But, by means of numerical integrations or dynamical maps it is easy to show that, due to the disruptive close encounters with Jupiter, the width for stable librations is approximately 0.17 au which is  the value predicted by our model discarding close encounters to less than 3$R_H$. 
Nevertheless,
in order
to reproduce the widths deduced from the dynamical map of figure 7 in
\cite{2019Icar..317..121G} corresponding to resonance 3:1 with Jupiter 
for $i=90^{\circ}$, at very high eccentricities we had to allow close encounters even to less than 1$R_H$ in our algorithm, probably because of the large orbital inclination but also because this resonance is extremely strong and can overcome such close encounters. On the other hand, for $e<0.7$ the match between the dynamical map and our widths for that resonance is almost perfect in any circumstances.
We also reproduced correctly the results shown in figure 8 of \cite{2019Icar..317..121G} because it is a configuration without close encounters ($i=\omega=90^{\circ}$), and the results of figure 17 from \cite{2019Icar..317..121G} allowing encounters as close as  1$R_H$.
Then the criterion of 3$R_H$ is a general rule but in particular cases it can be revised.

A limitation of our model is that
it does not take into account the law of structure \citep{1988AJ.....96..400F} that relates  $a_0$ with $e$, but this effect is restricted to  first order resonances at very low eccentricities.
It is originated in the non-negligible value of $\dot{\varpi}$,  which is typical of near zero  eccentricity orbits and that we have ignored assuming  constant $\varpi$ in Eq. (\ref{crit}).
In conclusion, we can say that this model allows to obtain a good approximation of the fundamental properties of MMRs in general in the Solar System. 
In the next section, applying this model  we will present a study of the resonant structure beyond Neptune in three dimensional space, that means, including the orbital inclination of the resonant objects. We will also show the relevance of the argument of the perihelion, $\omega$, for spatial resonances.

\section{Properties of the spatial MMRs beyond Neptune}
\label{proper}

\subsection{Libration centers, periods and widths for stable librations}

It is known that for the planar case all resonances except resonances
1:$k$ have libration centers strictly at  $\sigma_0=0^{\circ}$ or   $\sigma_0=180^{\circ}$. Resonances 1:$k$ instead have asymmetric librations, that means libration centers whose positions depend on the orbital eccentricity.
Applying our method to the spatial case we found that all resonances can have libration centers widely distributed in the interval $0^{\circ} \leq \sigma_0 \leq 360^{\circ}$ and that for a specific resonance, $\sigma_0$  depends on the set $(e,i,\omega)$.

Nevertheless, we have found a very particular situation when
 $\omega=N\times 90^{\circ}$ being $N$ an integer: for all spatial resonances the equilibrium points present a symmetry with respect to $\sigma=180^{\circ}$. 
 In this situation, resonances 1:$k$ exhibit a wide variation in the location of the equilibrium points but preserving the symmetry with respect to $\sigma=180^{\circ}$, while for all other resonances the 
 equilibrium points are strictly at $\sigma=$ 0$^{\circ}$ or 180$^{\circ}$.  
On the contrary,
for  $\omega\neq N\times 90^{\circ}$ the symmetry is destroyed for all resonances and the equilibrium points can be located in all the interval of $\sigma$ between 0$^{\circ}$ and 360$^{\circ}$.

\begin{figure}
	\includegraphics[width=1\textwidth]{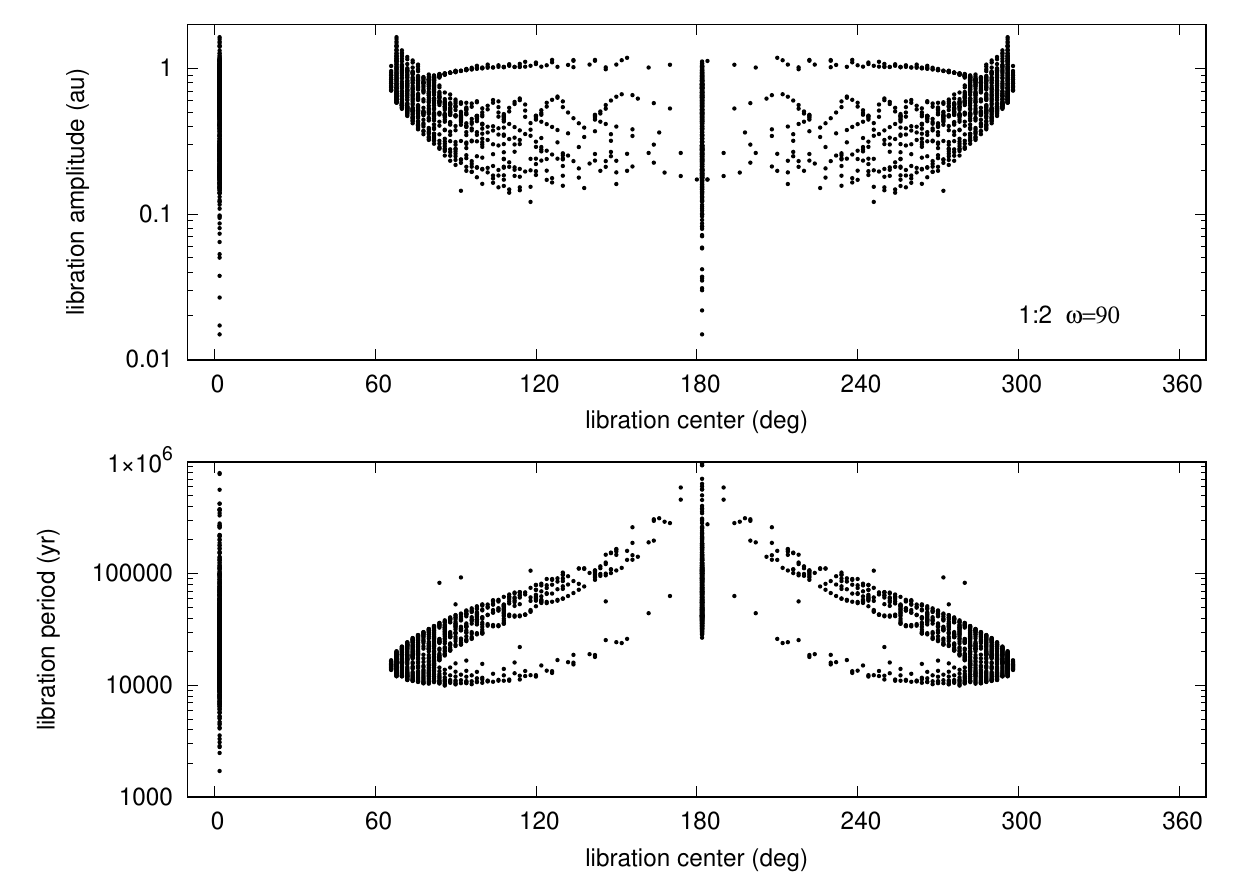}
	\caption{Resonance 1:2. Libration centers, periods (bottom) and maximum widths (top) for stable librations obtained when varying $e$ between 0.02 and 0.96 in steps of 0.02 and $i$ between 0$^{\circ}$ and 180$^{\circ}$ in steps of 5$^{\circ}$ assuming $\omega=90^{\circ}$. The symmetry with respect to $\sigma=180^{\circ}$ is preserved. All resonances 1:$k$ exhibit a similar behavior  when $\omega=N\times 90^{\circ}$.} 
	\label{1to2w90lib}     
\end{figure}

\begin{figure}
	\includegraphics[width=1\textwidth]{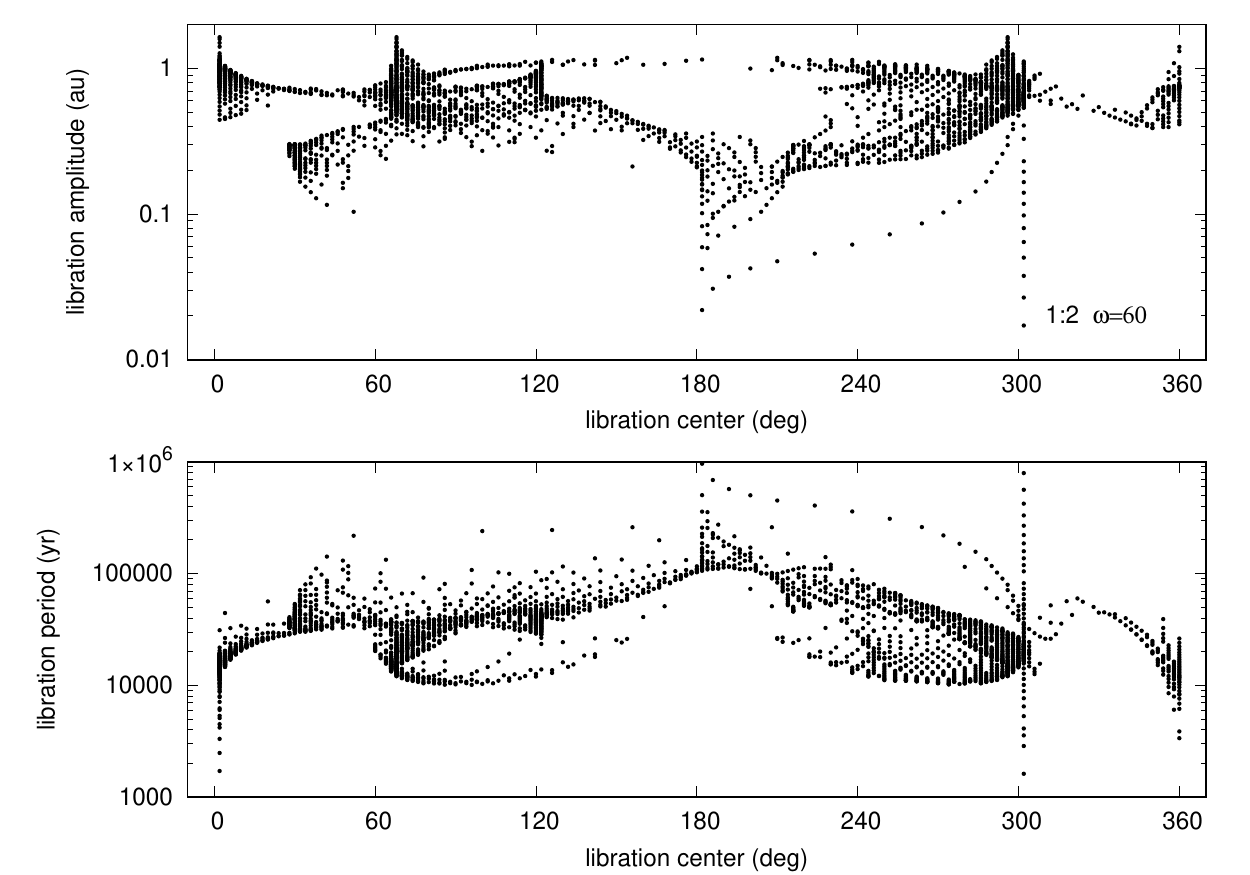}
	\caption{Resonance 1:2. Same as Fig.  \ref{1to2w90lib} but assuming $\omega=60^{\circ}$. The symmetry is destroyed.}
	\label{1to2w60lib}     
\end{figure}

\begin{figure}
	\includegraphics[width=1\textwidth]{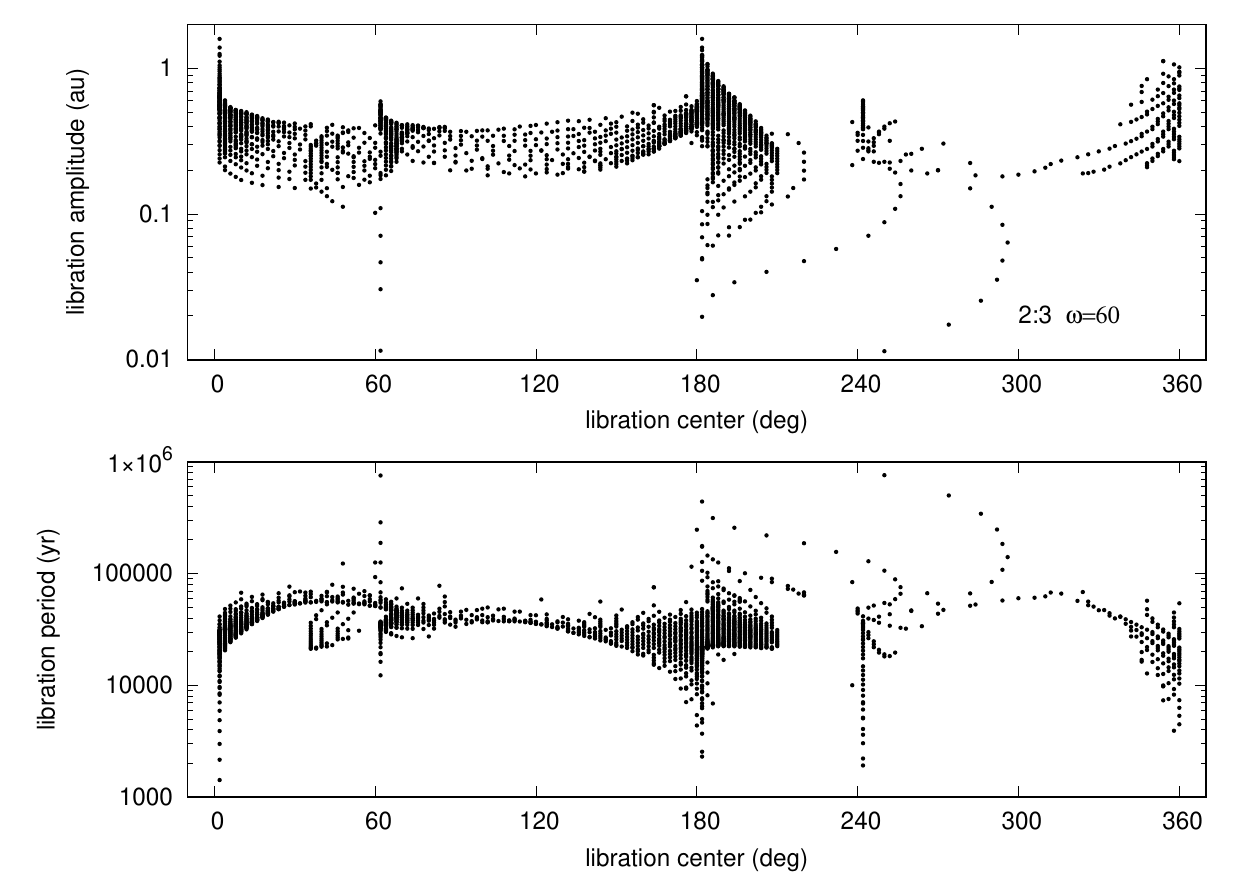}
	\caption{Same as Fig.  \ref{1to2w60lib} but for resonance 2:3   ($\omega=60^{\circ}$). }
	\label{2to3w60lib}     
\end{figure}

We illustrate these properties with some examples. Fig.  \ref{1to2w90lib} shows all the libration centers, periods and widths we have found for resonance 1:2 when assuming $\omega=90^{\circ}$ and varying the eccentricity between 0.02 and 0.96 in steps of 0.02 and the inclination between 0$^{\circ}$ and 180$^{\circ}$ in steps of 5$^{\circ}$. 
Depending on the values of $(e,i)$ sometimes we found one or two or three stable libration centers and we plotted all them.
 There is a perfect symmetry with respect to $\sigma=180^{\circ}$ and similar situations occur when $\omega=N\times 90^{\circ}$.
Fig.  \ref{1to2w60lib} shows the same resonance but imposing $\omega=60^{\circ}$, the symmetry is destroyed and the stable libration centers can be located in all interval between 0$^{\circ}$ and 360$^{\circ}$ depending on $(e,i)$. 
Fig.  \ref{2to3w60lib} shows the case of the resonance 2:3 imposing  $\omega=60^{\circ}$. The libration center can be located anywhere in the interval between 0$^{\circ}$ and 360$^{\circ}$, but taking $\omega=N\times 90^{\circ}$ we obtain libration centers exclusively at 0$^{\circ}$ or 180$^{\circ}$ for this resonance. Then, $\omega$ is a crucial parameter for the location of the libration centers in the spatial case.
Even more, we have found that the number, location, stability of the equilibrium points as well as the topology of $\mathcal{H}(a,\sigma)$ depend on the set $(e,i,\omega)$.

\begin{figure}
	\includegraphics[width=0.8\textwidth]{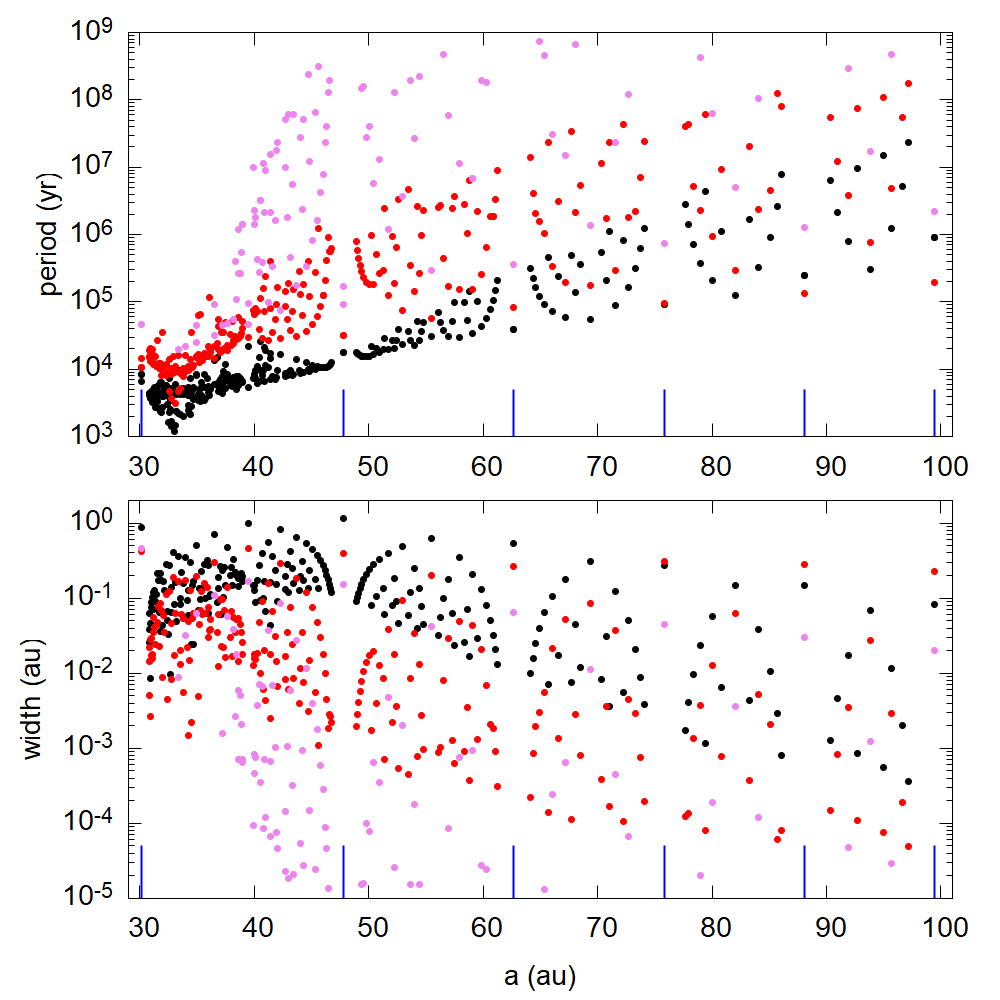}
	\caption{Libration periods of the small amplitude oscillations (top) and maximum widths (bottom) for all 234 resonances with $k_p,k\le 30$ located between 30 and 100 au. Calculations correspond to orbits with $e=0.3, \omega=90^{\circ}$ and for each resonance three different orbital inclinations were considered:  10$^{\circ}$, 90$^{\circ}$ and 170$^{\circ}$  represented with  black, red and violet points respectively. From left to right, location of resonances 1:1, 1:2, 1:3, 1:4, 1:5 and 1:6 are represented by short blue vertical lines.}
	\label{lib30100}     
\end{figure}

To get an idea of the wide variety of resonance properties,
a global representation of libration periods and maximum widths for all resonances between 30 and 100 au verifying $k_p,k\le 30$ is shown in Fig.  \ref{lib30100}.
All calculations correspond to orbits with arbitrarily chosen $e=0.3$ and $\omega=90^{\circ}$, and for each resonance we show the results for three different inclinations: 10$^{\circ}$ with black points, 90$^{\circ}$ in red and 170$^{\circ}$ in violet. For some resonances more than one libration state is possible and all them were plotted. 
We note that several resonances for the case $i=170^{\circ}$ with high $k_p,k$ close to resonance 1:1 were computed with zero width because of instabilities generated by close encounters with Neptune. For this particular inclination the resonances are in general weaker with the notable exception of resonances 1:$k$. Polar resonances ($i=90^{\circ}$) are sometimes as strong as the resonances for $i=10^{\circ}$.  
We tested our predicted libration periods for $i=0^{\circ}$ with the ones given by \cite{Lan2019} in their Fig.  8b  and we have found a very good agreement. 
In our Fig.  \ref{lib30100} it is evident that, for a given interval in $a$, the larger the resonance width the shorter the libration period of the small amplitude librations, which is related to better stability.
Libration periods of several Myrs probably are not realistic because they are associated to weak resonances and also because
we are not considering the secular effects that are characteristic of the TNR.
Resonances 1:$k$ marked with a short vertical blue line in Fig.  \ref{lib30100} appear the strongest, widest and with the shortest libration periods, independently of their inclinations, when compared with their surrounding resonances. They dynamically dominate because of their larger strength and isolation and their dynamical relevance shapes
distribution of the points in Fig. \ref{lib30100}.

\subsection{The fragility of the spatial resonances}

\begin{figure}
	\includegraphics[width=1\textwidth]{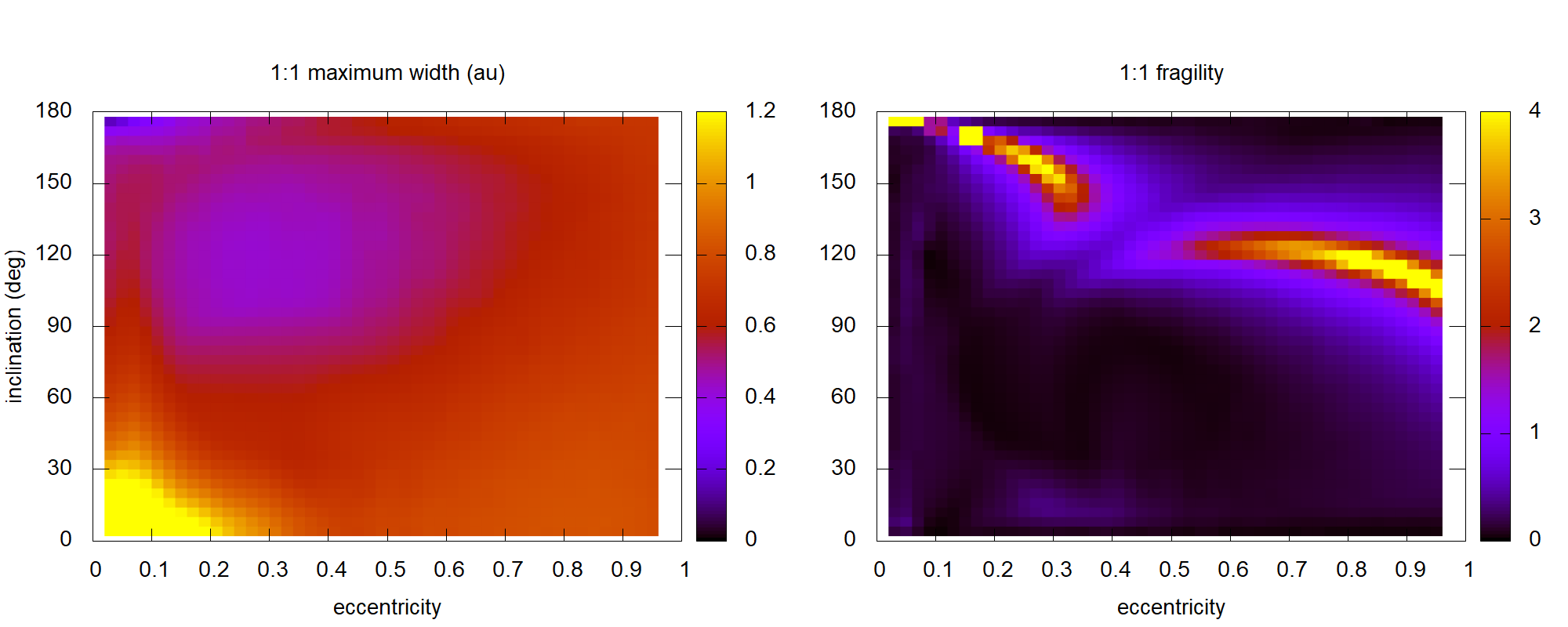}
	\caption{Resonance 1:1 at $a=30.1$ au. Left: maximum width in color code from 0 to 1.2 au obtained varying $\omega$. Right: fragility in color code from 0 to 4. Yellow regions represent very high fragility regions with $f\geq 4$.} 
	\label{1to1f}     
\end{figure}

\begin{figure}
	\includegraphics[width=1\textwidth]{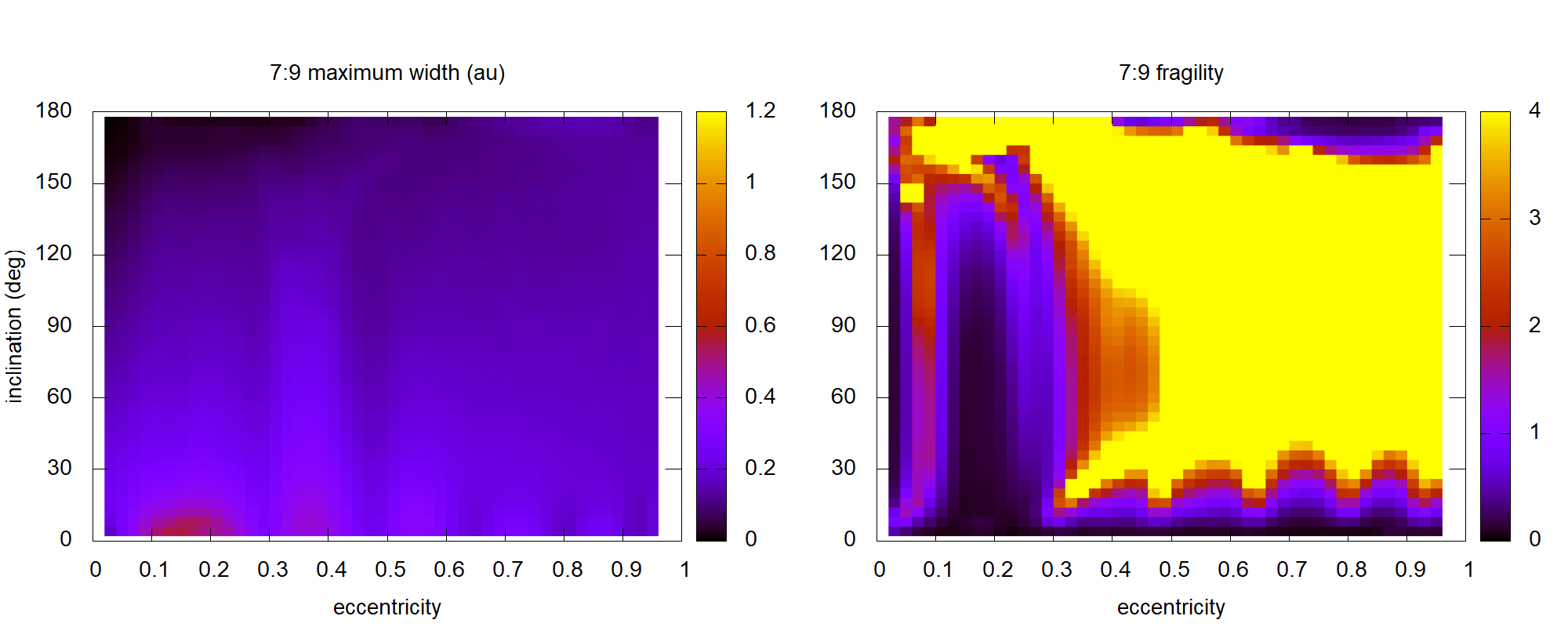}
	\caption{Same as Fig. \ref{1to1f} for  resonance 7:9 at $a=35.6$ au. Left: maximum width in au obtained varying $\omega$. Right: fragility.  }
	\label{7to9f}     
\end{figure}

\begin{figure}
	\includegraphics[width=1\textwidth]{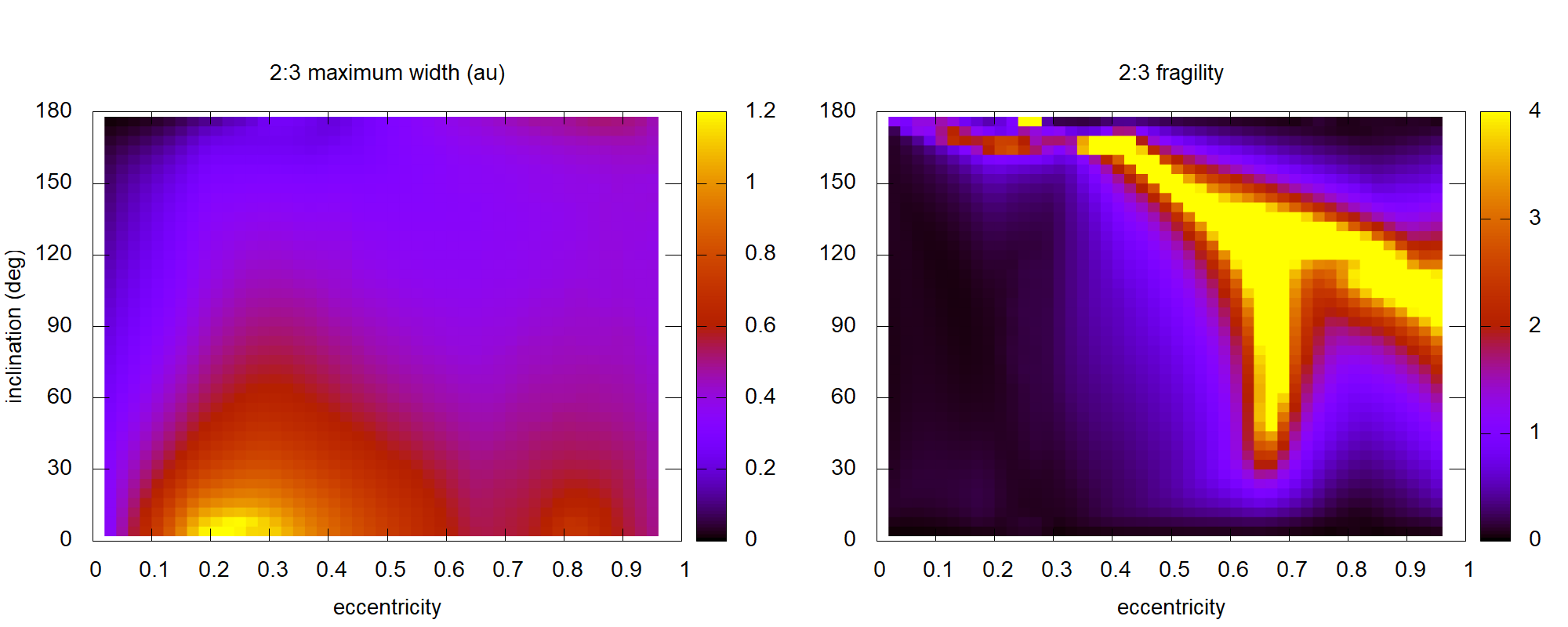}
	\caption{Same as Fig. \ref{1to1f} for   resonance 2:3 at $a=39.5$ au. Left: maximum width in au obtained varying $\omega$. Right: fragility.}
	\label{2to3f}     
\end{figure}

\begin{figure}
	\includegraphics[width=1\textwidth]{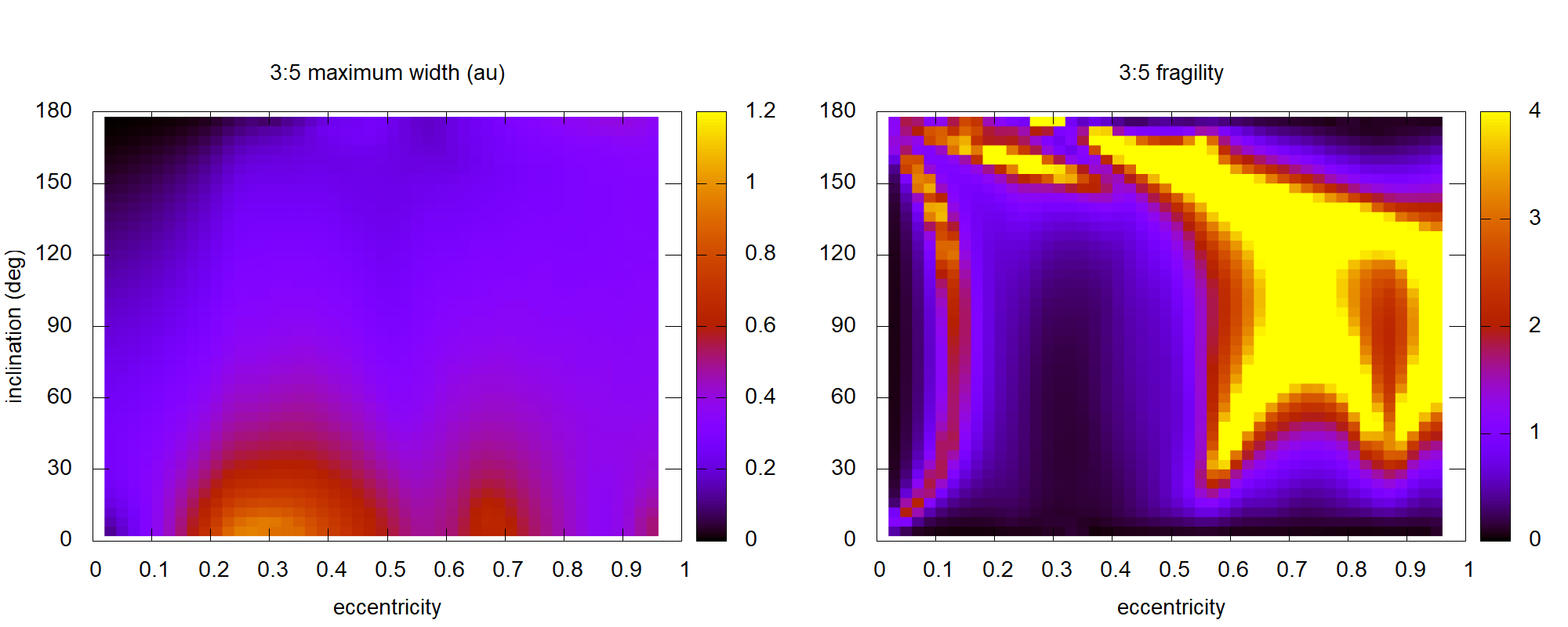}
	\caption{Same as Fig. \ref{1to1f} for   resonance 3:5 at $a=42.3$ au. Left: maximum width in au obtained varying $\omega$. Right: fragility.}
	\label{3to5f}     
\end{figure}

\begin{figure}
	\includegraphics[width=1\textwidth]{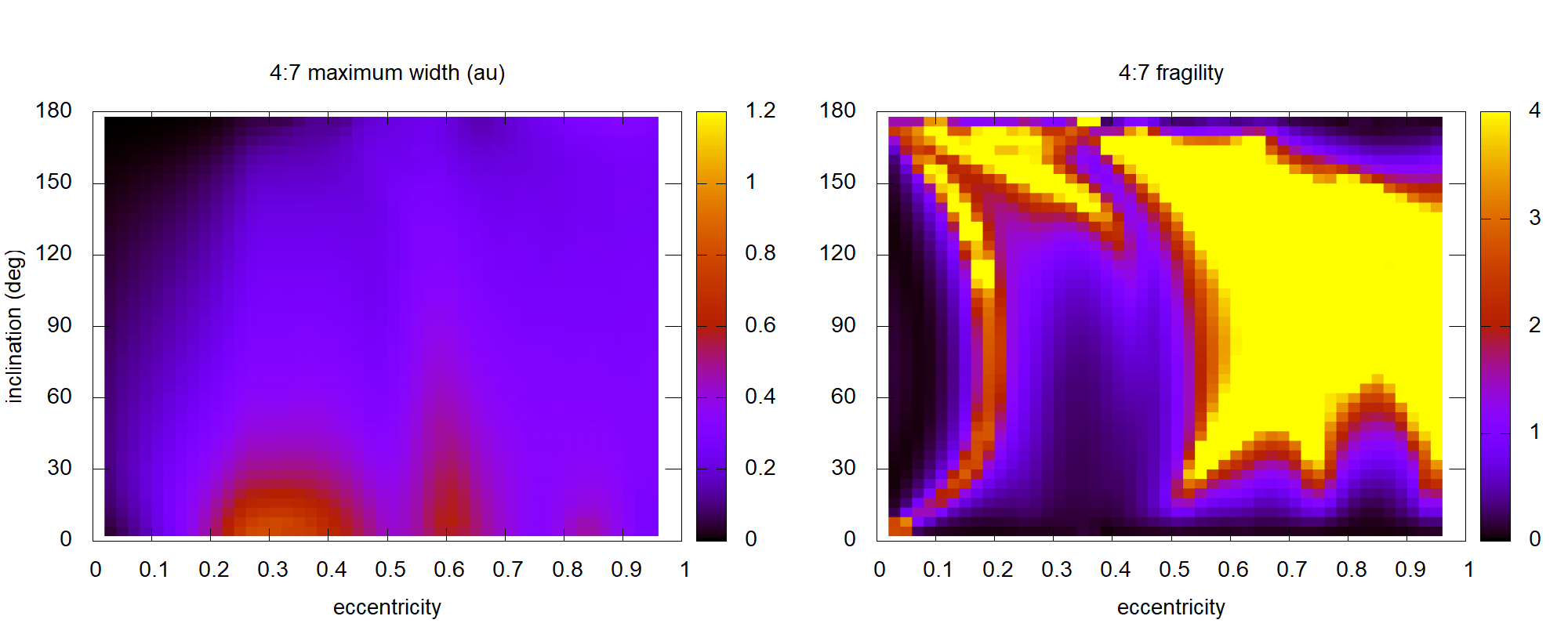}
	\caption{Same as Fig. \ref{1to1f} for   resonance 4:7 at $a=43.7$ au. Left: maximum width in au obtained varying $\omega$. Right: fragility.}
	\label{4to7f}     
\end{figure}

\begin{figure}
	\includegraphics[width=1\textwidth]{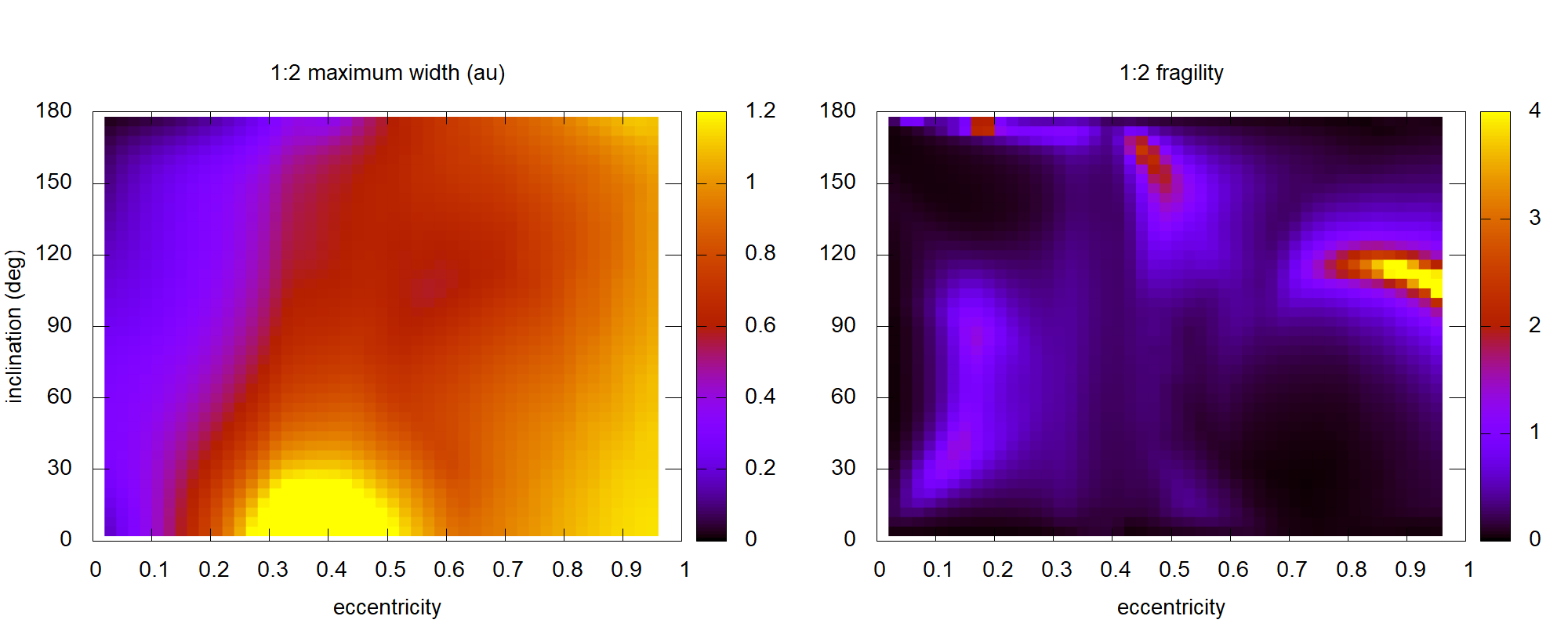}
	\caption{Same as Fig. \ref{1to1f} for   resonance 1:2 at $a=47.8$ au. Left: maximum width in au obtained varying $\omega$. Right: fragility.}
	\label{1to2f}     
\end{figure}

\begin{figure}
	\includegraphics[width=1\textwidth]{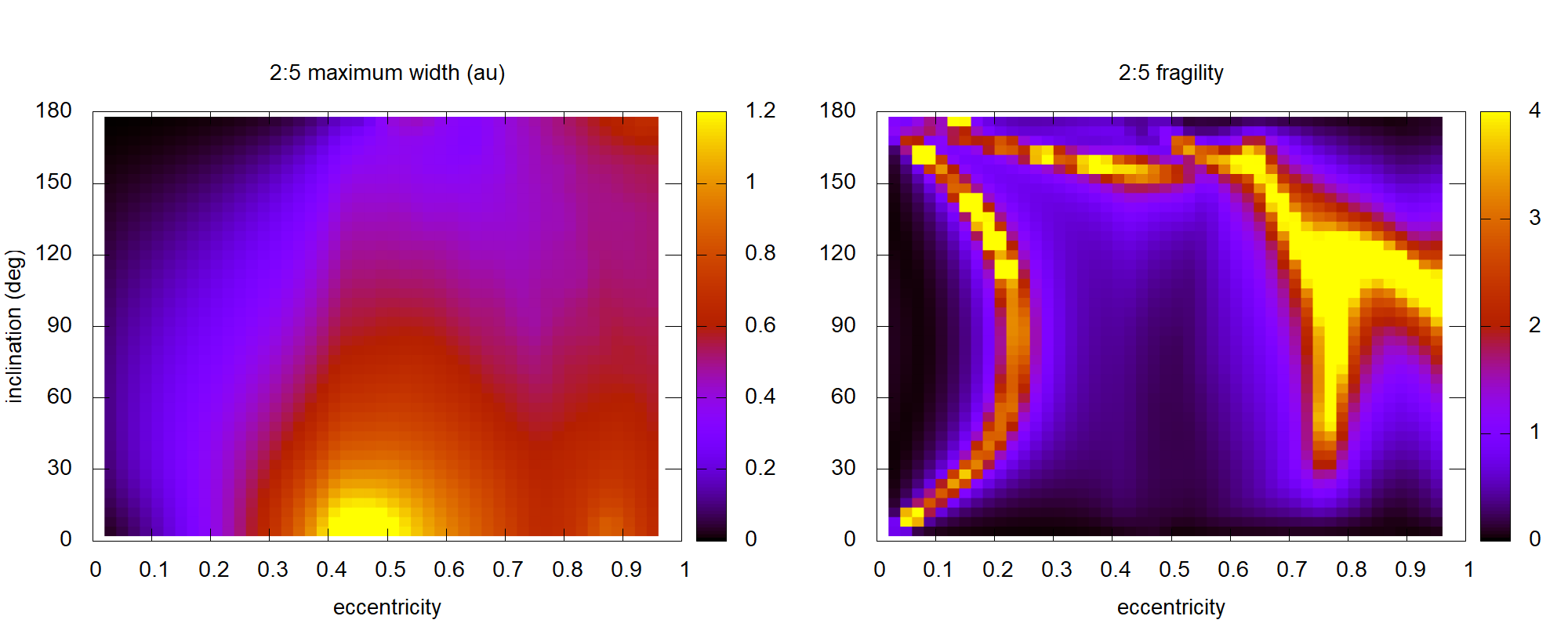}
	\caption{Same as Fig. \ref{1to1f} for   resonance 2:5 at $a=55.5$ au. Left: maximum width in au obtained varying $\omega$. Right: fragility.}
	\label{2to5f}     
\end{figure}

\begin{figure}
	\includegraphics[width=1\textwidth]{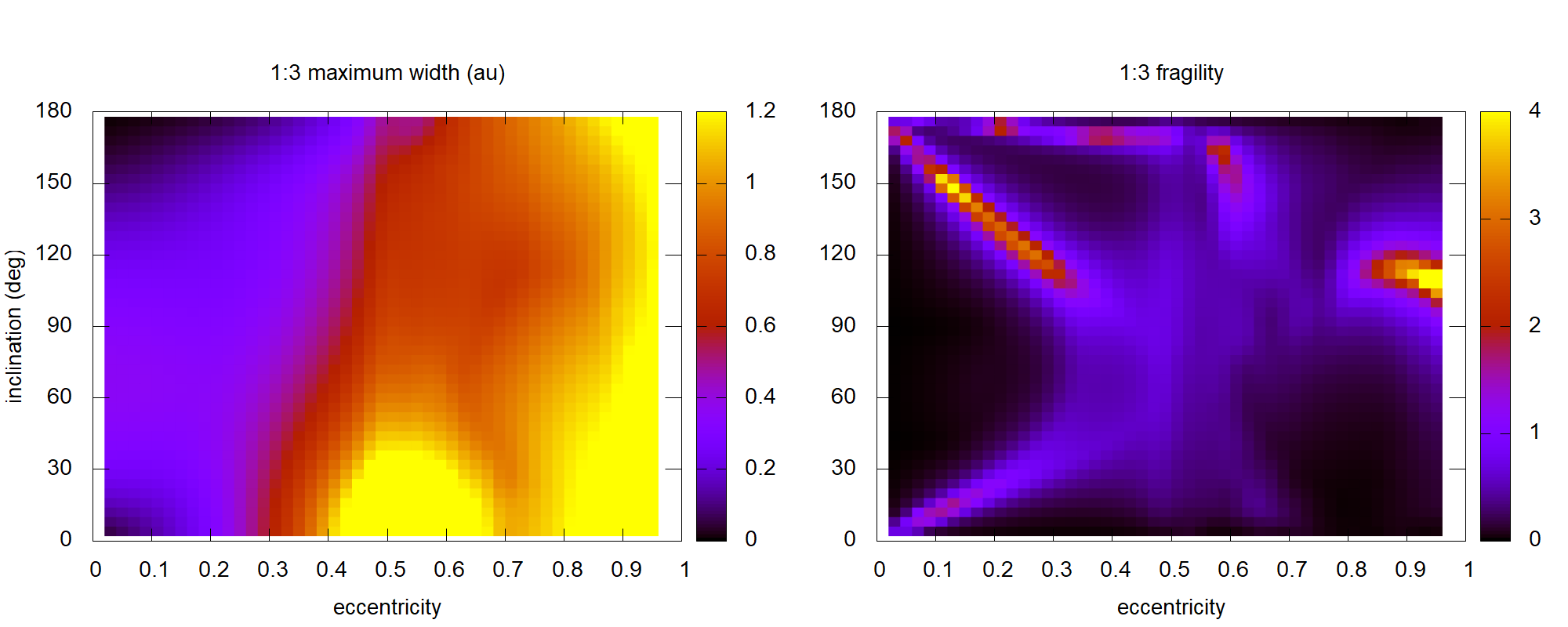}
	\caption{Same as Fig. \ref{1to1f} for   resonance 1:3 at $a=62.6$ au. Left: maximum width in au obtained varying $\omega$. Right: fragility.}
	\label{1to3f}     
\end{figure}

In the planar model the strength and width of a given resonance depend just on the orbital eccentricity. In the spatial case they depend on the set of parameters $(e,i,\omega)$. While the orbit is librating, $(e,i)$ experience small amplitude oscillations but $\omega$ generally circulates or shows large variations in longer timescales. The variations in $\omega$ generate changes in the resonance width and topology. If these changes are small, the resonance will not be affected but if they are large eventually the resonance could become weak and the resonant motion may break down. Then, for the spatial resonances we introduce a new concept, that we will call resonance \textit{fragility}, and that we define it as the dimensionless parameter
\begin{equation}\label{eqfrag}
f(e,i)=(\Delta a_{max}-\Delta a_{min})/\Delta a_{min} 
\end{equation}
where $\Delta a_{max},\Delta a_{min}$ are the maximum and minimum total resonance widths of stable librations obtained when varying $\omega$ for fixed $(e,i)$.
A fragility equal to 0 means that there is no fragility, the resonance's width is invariable with $\omega$ and the resonance can be considered stable in the sense that changes in its properties cannot be expected. A resonance with fragility $f$ can change its width by a factor of $f+1$, so it is an indication of instability. 
Then, for a given resonance,  $f$ is a function depending on $(e,i)$ and we could have regions of the plane $(e,i)$ with high fragility. These regions indicate the values $(e,i)$ for which the resonant motion is more vulnerable, or fragile, and
the least probable regions capable of sustain a resonant population for long time scales.
In Figs. \ref{1to1f} to \ref{1to3f} we illustrate with some resonances that have been considered since the study of the dynamics of the TNR began ordered in increasing values of semimajor axes. In the left panels, on a grid of $(e,i)$, we show the maximum widths of stable librations in au calculated when varying  $\omega$ from 0$^{\circ}$ to 90$^{\circ}$ ($\Delta \mathcal{R}$ is $\pi$-periodic in $\omega$ and symmetric respect to $\omega=90^{\circ}$) in steps of 5$^{\circ}$  and in the right panels the corresponding fragility $f$ according to Eq. (\ref{eqfrag}). The same scale for $\Delta a_{max}$ was used in all figures in order to an easy comparison between them and the same scale was used in all figures for $f$.

For low inclination orbits, say $i<10^{\circ}$, the fragility is in general very low but for higher inclinations some resonances show increasing fragility. We have found that high fragility is associated with changes in the stability of the equilibrium points. 
Although high fragility is associated with lower resonance widths we have found that the fragility mostly depends on the type of the resonance.
 Resonances
1:1, 1:2 and 1:3 show larger widths in au and these are the resonances which exhibit lower fragility. More precisely, examining the figures it is evident that 
for a generic resonance $k_p$:$k$
the larger the value of $k_p$ the larger the fragility of the resonance being 7:9, 4:7 and 3:5 the most fragile of the resonances shown and in that order.
Then, resonances 1:$k$ ($k_p=1$) are the widest and the more robust, followed by 2:$k$  ($k_p=2$).
The polar resonant object 471325 (2011 KT19) is evolving in the region $.28<e<.48$ and $i\sim 112^{\circ}$ of resonance 7:9  \citep{2017MNRAS.472L...1M}  where according to Fig.  \ref{7to9f}  the fragility is large, more precisely it varies from 2 to 6. It is a very fragile region and in fact the object leaves the resonance after some Myrs. We have not studied experimentally the effect of $f$ in a population of resonant objects, but 
	if we imagine that $N$ resonant objects are uniformly distributed along the domain of the resonance,
	we can guess that a resonance with fragility $f$ could loss its members until
	reduced to  $N/(f+1)$. On the other hand, an object observed evolving inside a region of high fragility must be a survivor of an originally larger population. That could be the case for  471325 (2011 KT19).

\cite{2012Icar..220..392G} and \cite{2016CeMDA.126..369S,2017CeMDA.127..477S} studied the long term dynamical evolution of resonant motions due to the Lidov-Kozai mechanism where large changes in $(e,i)$ take place along with variations of $\omega$. In these cases, the resonances also have large variations in their behavior and strength, including notable changes in their topology. Our definition of fragility only takes into account the variations generated by $\omega$, independently of the secular mechanism affecting  $(e,i)$ of the objects. Fragility is an intrinsic property of the resonance 
and is defined by the values of $(e,i)$. Nevertheless, if we know the long term time evolution of $(e,i)$, we can follow the evolution of the resonant motion in terms of width and fragility. For example, if the object is evolving towards a region of  high fragility in the plane $(e,i)$ the resonance could break, and on the other hand, if it is evolving towards  a region of low fragility the resonance will be guaranteed.

\subsection{Atlas of resonances from 30 to 100 au}

\begin{figure}
	\includegraphics[width=1\textwidth]{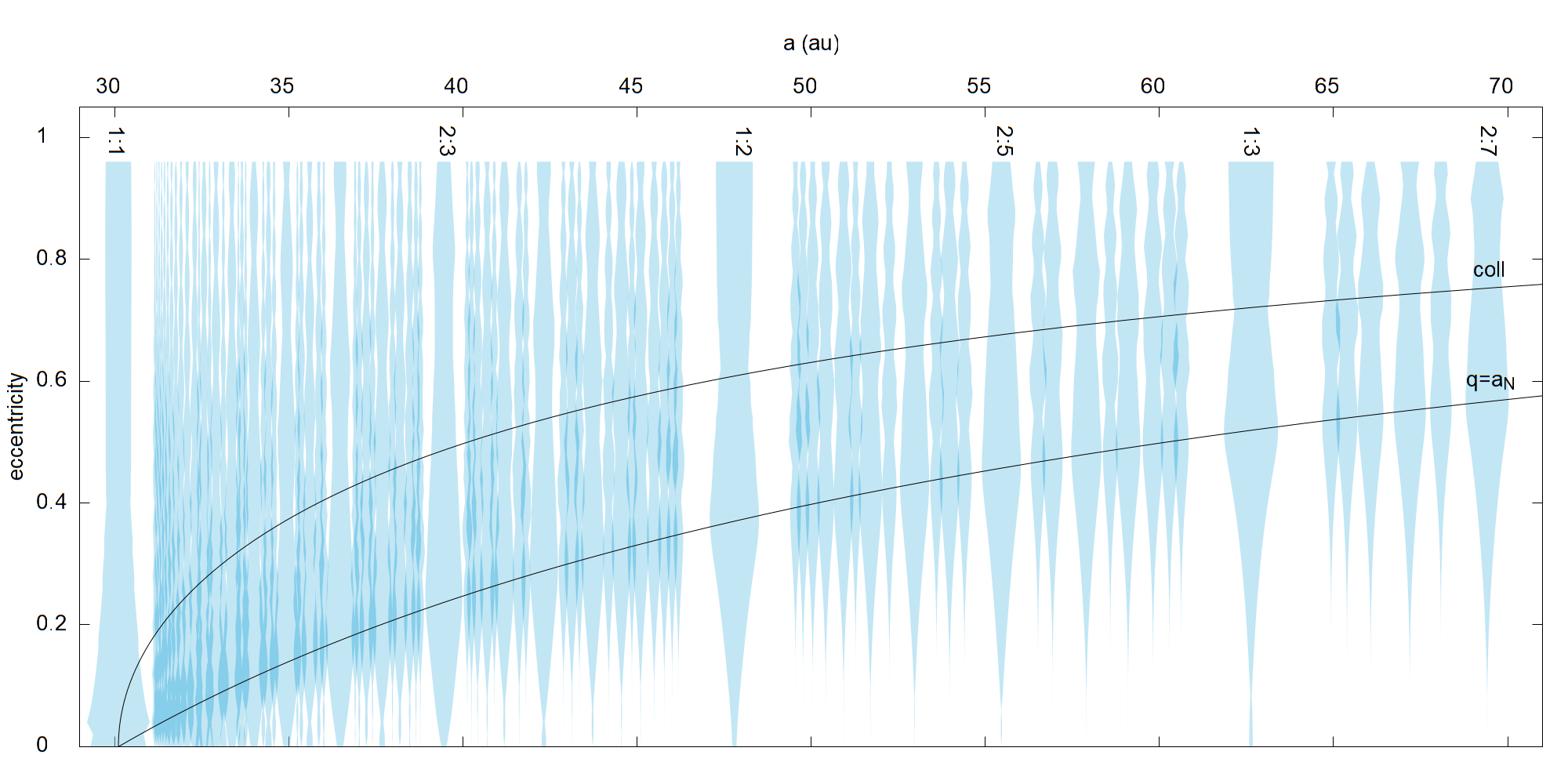}
	\caption{Resonance widths for all the 93 resonances with Neptune between 30 and 70 au verifying $k\leq 20$ and  $k_p\leq 20$ calculated with $i=10^{\circ}$  and $\omega=90^{\circ}$. Collision curve from Eq. (\ref{col}) and the curve corresponding to $q=a_N$ are shown. Resonances 1:$k$ are strong and isolated from their neighbors.}
	\label{atlasi10}     
\end{figure}

To have a general panorama of the resonances beyond Neptune we show in Fig.  \ref{atlasi10} the classic picture of resonance's widths as function of the eccentricity but calculated for $i=10^{\circ}$ and $\omega=90^{\circ}$ for all resonances with Neptune with 
 $k_p\leq 20$ and $k\leq 20$ . The darkest regions are due to the superposition of resonances. 
In the planar theory, when the perihelion $q=a(1-e)$ verify $q\leq a_N$ the intersection of orbits is unavoidable and collisions take place unless the critical angle is limited to safe values. In the spatial case instead, as is the case of Fig.  \ref{atlasi10}, if we assume that the planet has a zero eccentricity and zero inclination orbit, the condition for intersection of orbits is given by 
\begin{equation}\label{col}
\frac{a(1-e^2)}{1\pm e\cos \omega} = a_p
\end{equation}
where $a_p$ is the semimajor axis of the planet, Neptune in this case ($a_p=a_N$).
This collision curve, $e(a)$ for $\omega=90^{\circ}$, and the curve given by $q = a_N$ (which defines the \textit{critical} eccentricity $e_c$) are shown in Fig.  \ref{atlasi10} . The last one is associated to regions in the plane $(a,e)$ where resonances are wider and, on the contrary, the collision curve is associated to regions where resonances shrink due to the restrictions to the values of $\sigma$ imposed by the close encounters with Neptune.
Note the domain and isolation of resonances 1:$k$ (including 1:1) and in a lesser extent resonances 2:$k$. Note also the superposition of resonances for $a<34$ au which it is known to be a chaotic region \citep{1993ApJ...406L..35L}. Our widths are in very good agreement with the results presented in Fig.  1 by \cite{Lan2019} for the planar case.

\begin{figure}
	\includegraphics[width=1\textwidth]{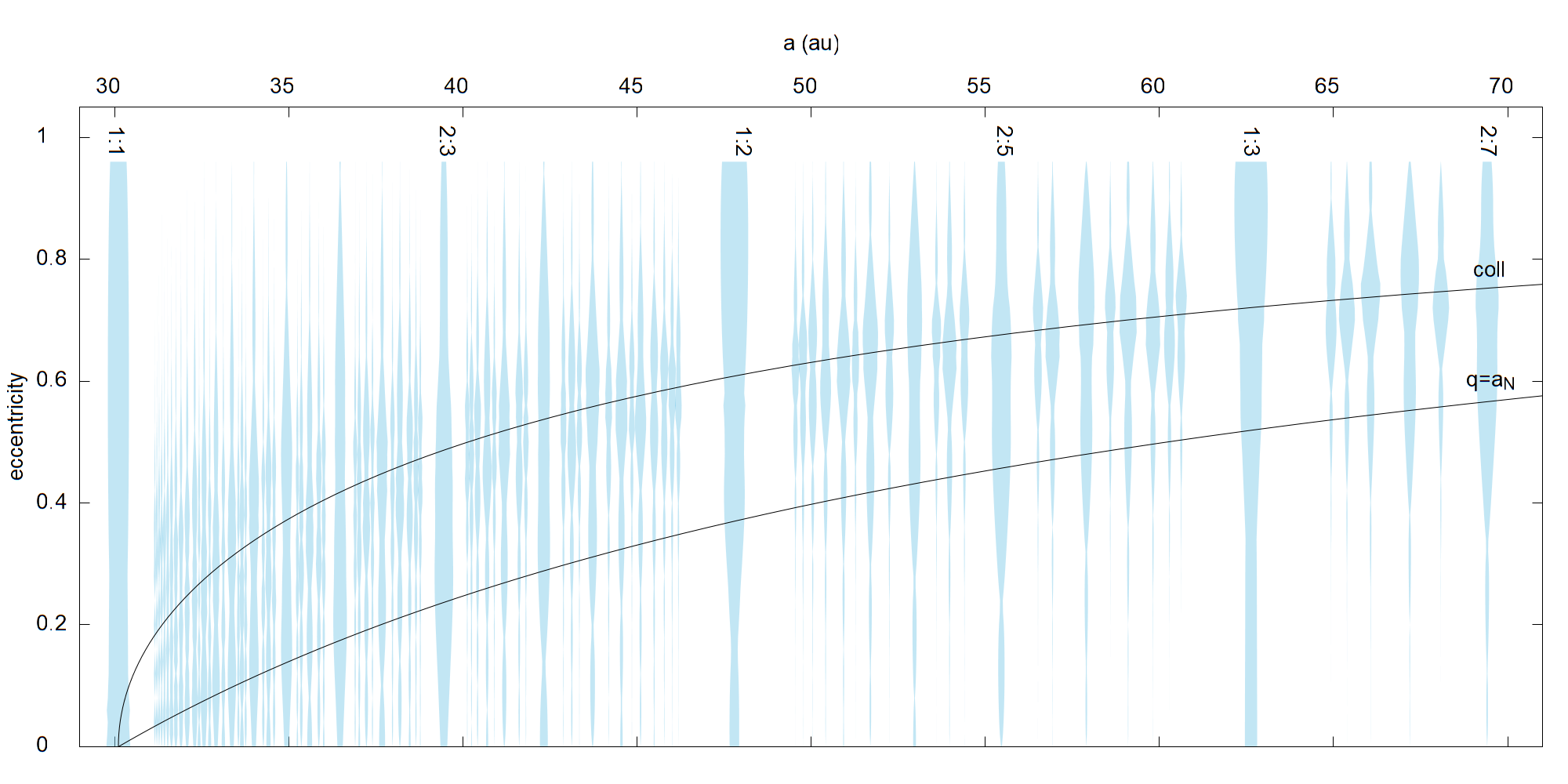}
	\caption{Same as Fig.  \ref{atlasi10} but calculated for  $i=70^{\circ}$ and $\omega=90^{\circ}$. Resonances 1:$k$ are still strong and isolated. }
	\label{atlasi70}     
\end{figure}

\begin{figure}
	\includegraphics[width=1\textwidth]{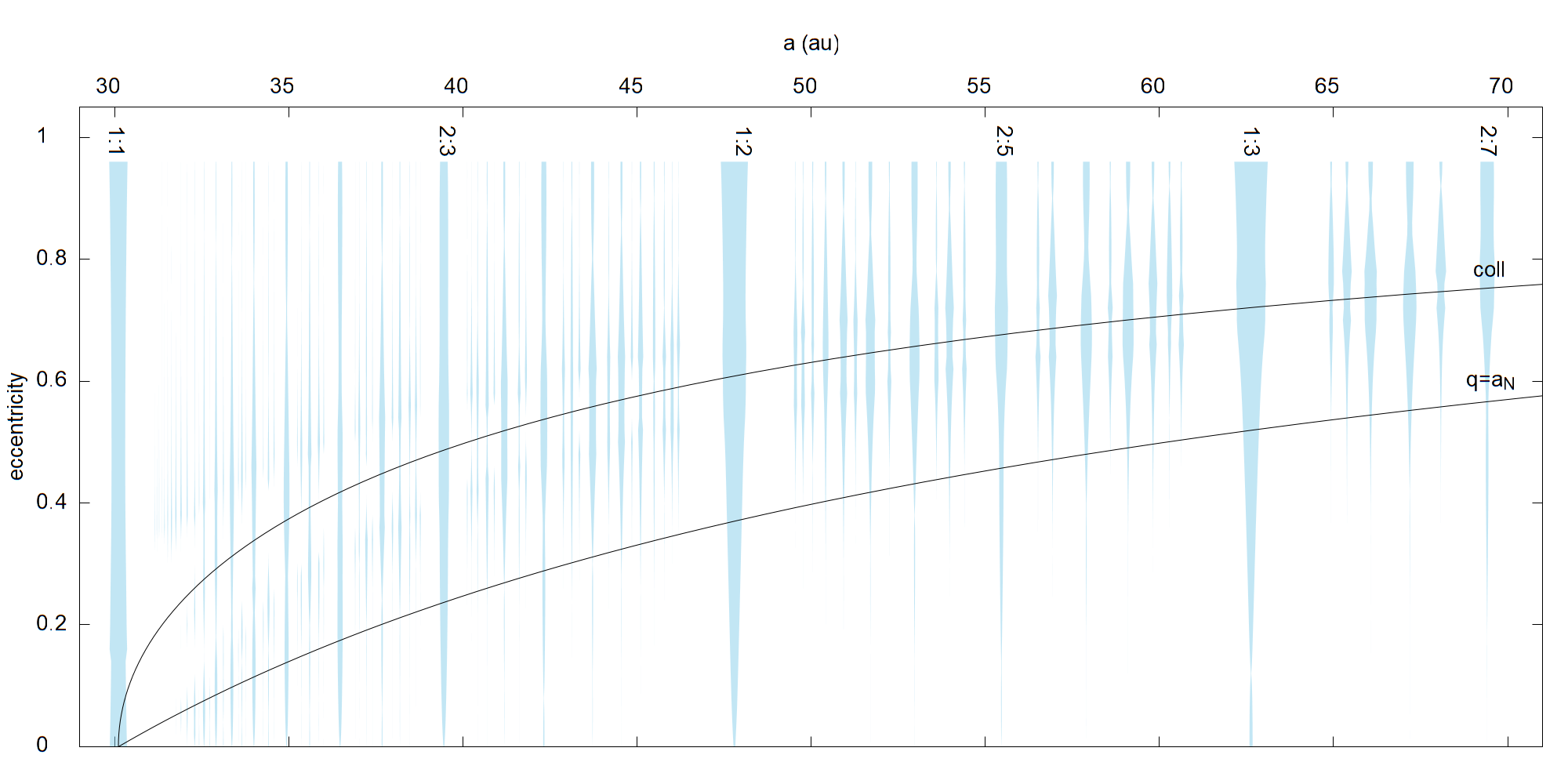}
	\caption{Same as Fig.  \ref{atlasi10} but calculated for  $i=150^{\circ}$ and $\omega=90^{\circ}$. Resonances 1:$k$ and 2:$k$ dominate. }
	\label{atlasi150}     
\end{figure}

\begin{figure}
	\includegraphics[width=0.7\textwidth]{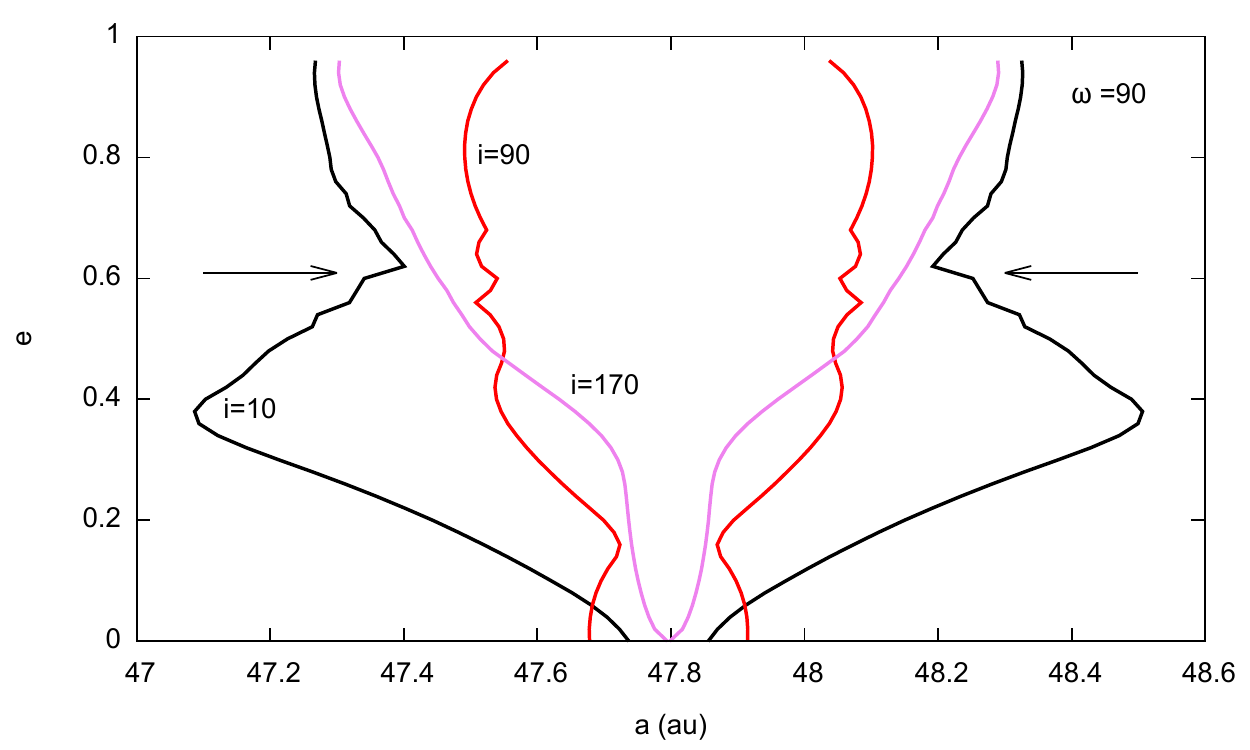}
	\caption{Resonance widths for resonance 1:2 computed for three different inclinations $i=$ 10$^{\circ}$, 90$^{\circ}$ and 170$^{\circ}$ assuming   $\omega=90^{\circ}$.  The collision eccentricity with Neptune deduced from Eq. (\ref{col})  is indicated with arrows.}
	\label{1to2three}     
\end{figure}

To illustrate the inclination effect we show in Fig.  \ref{atlasi70} the same resonances of Fig.  \ref{atlasi10} but calculated for $i=70^{\circ}$. For very low eccentricities they are wider than in the case of $i=10^{\circ}$ but globally, considering all the interval of eccentricities, they become narrower avoiding their superposition. 
Resonances 1:$k$ and 2:$k$ continue to dominate.
In Fig.  \ref{atlasi150} we show the panorama for $i=150^{\circ}$. In this case, resonances 1:$k$ and 2:$k$ persist and the others present some predominance only close to the collision curve.
We illustrate with more detail
the effect of the orbital inclination on the resonance width  in Fig.  \ref{1to2three} for the case of resonance 1:2 for three extreme values of inclination. The orbital inclination (and also $\omega$) changes completely the resonance domain in $(a,e)$.

A very interesting effect appears for high inclination resonances that can be observed in Figs. \ref{atlasi70} and \ref{atlasi150}: they are wider close to the collision curve. We checked this behavior with dynamical maps confirming the predictions of our model. Then, high inclination non resonant objects in collision routes could be trapped by weak resonances which are abnormally wide for that particular eccentricities.

\begin{figure}
	\includegraphics[width=1\textwidth]{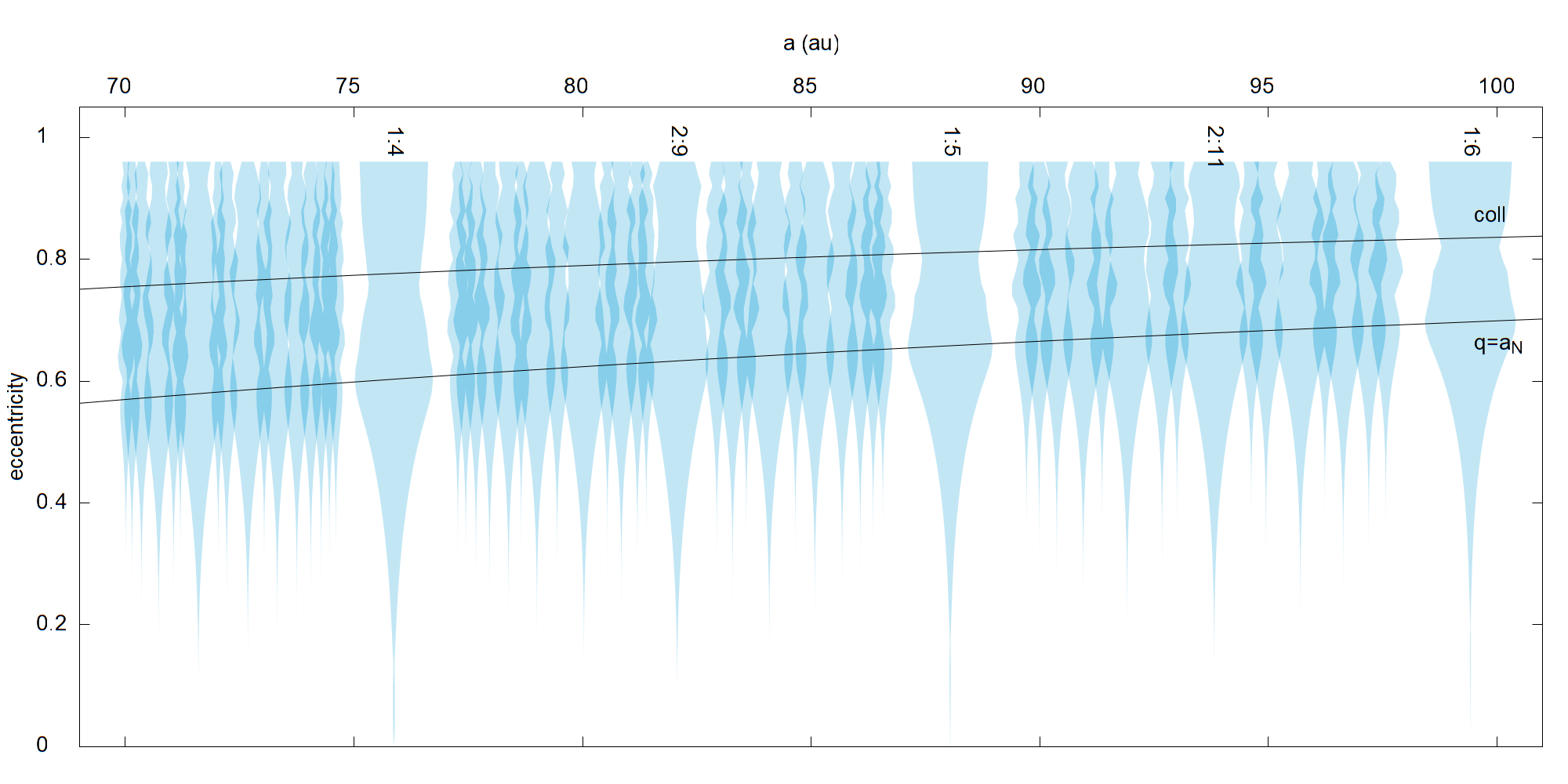}
	\caption{Resonance widths for  the 60 resonances with Neptune between 70 and 100 au verifying $k\leq 40$ and  $k_p\leq 40$ calculated for $i=10^{\circ}$  and $\omega=90^{\circ}$.  }
	\label{atlasi10b}     
\end{figure}

\begin{figure}
	\includegraphics[width=1\textwidth]{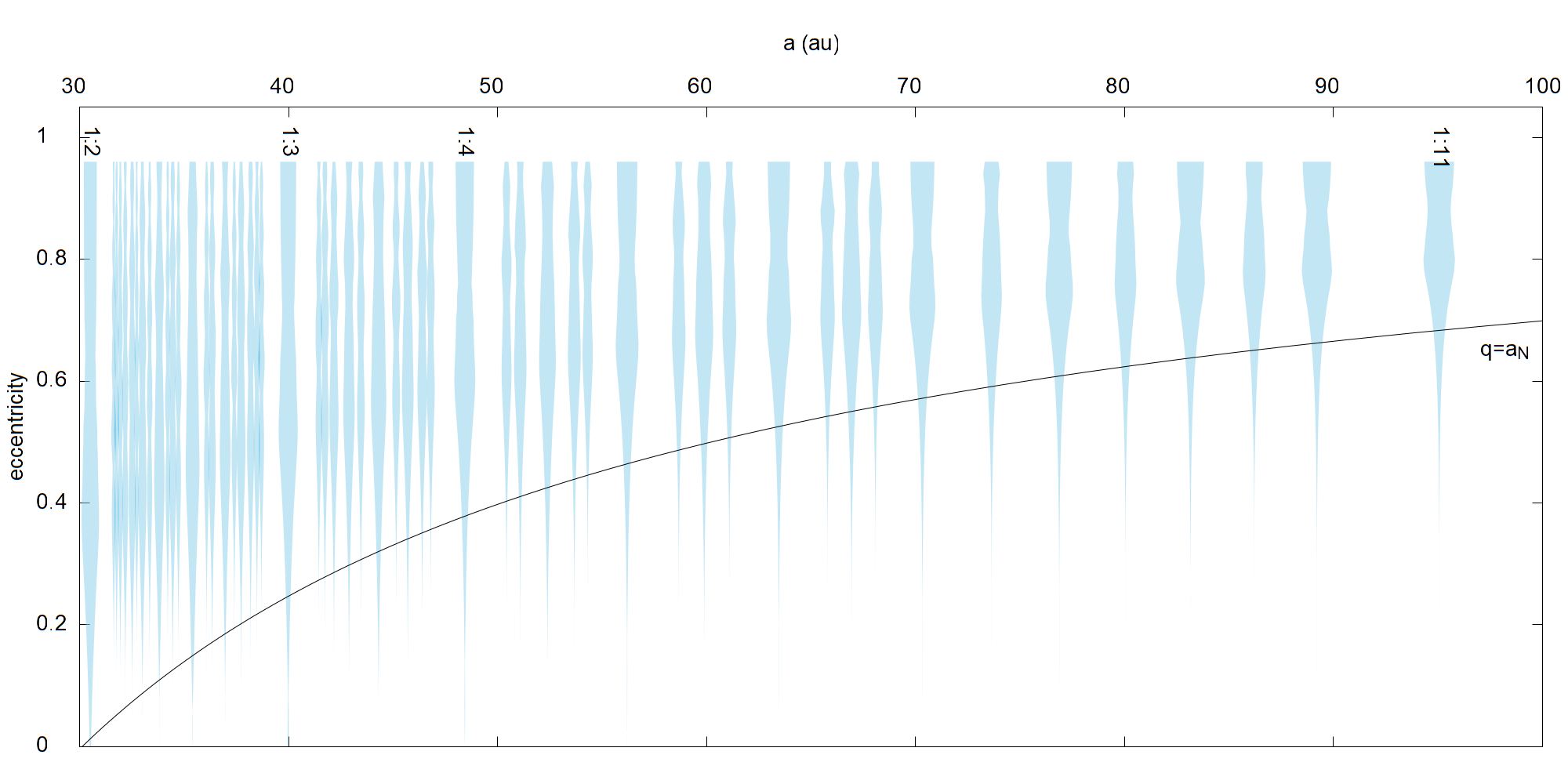}
	\caption{Resonances with Uranus calculated for $i=10^{\circ}$ and $\omega=90^{\circ}$. Only orbits with eccentricity below the curve $q=a_N$ are stable.}
	\label{uranus}     
\end{figure}

Going further,
Fig.  \ref{atlasi10b} shows the distribution of resonances between 70 and 100 au calculated as in Fig.  \ref{atlasi10} but up to $k_p= 40$ and  $k= 40$. Again, resonances 1:$k$ show up strong and isolated. Finally, resonances with Uranus are shown in Fig.  \ref{uranus}. In this case only orbits below the curve $q=a_N$ can survive in long timescales. At these low eccentricities, resonances with Uranus are weak so not very much objects can evolve in these resonances. Nevertheless, their imprint can appear in dynamical maps for example, as we will show later.

\subsection{On the existence of high $k_p$:$k$ resonances}

\begin{figure}
	\includegraphics[width=1\textwidth]{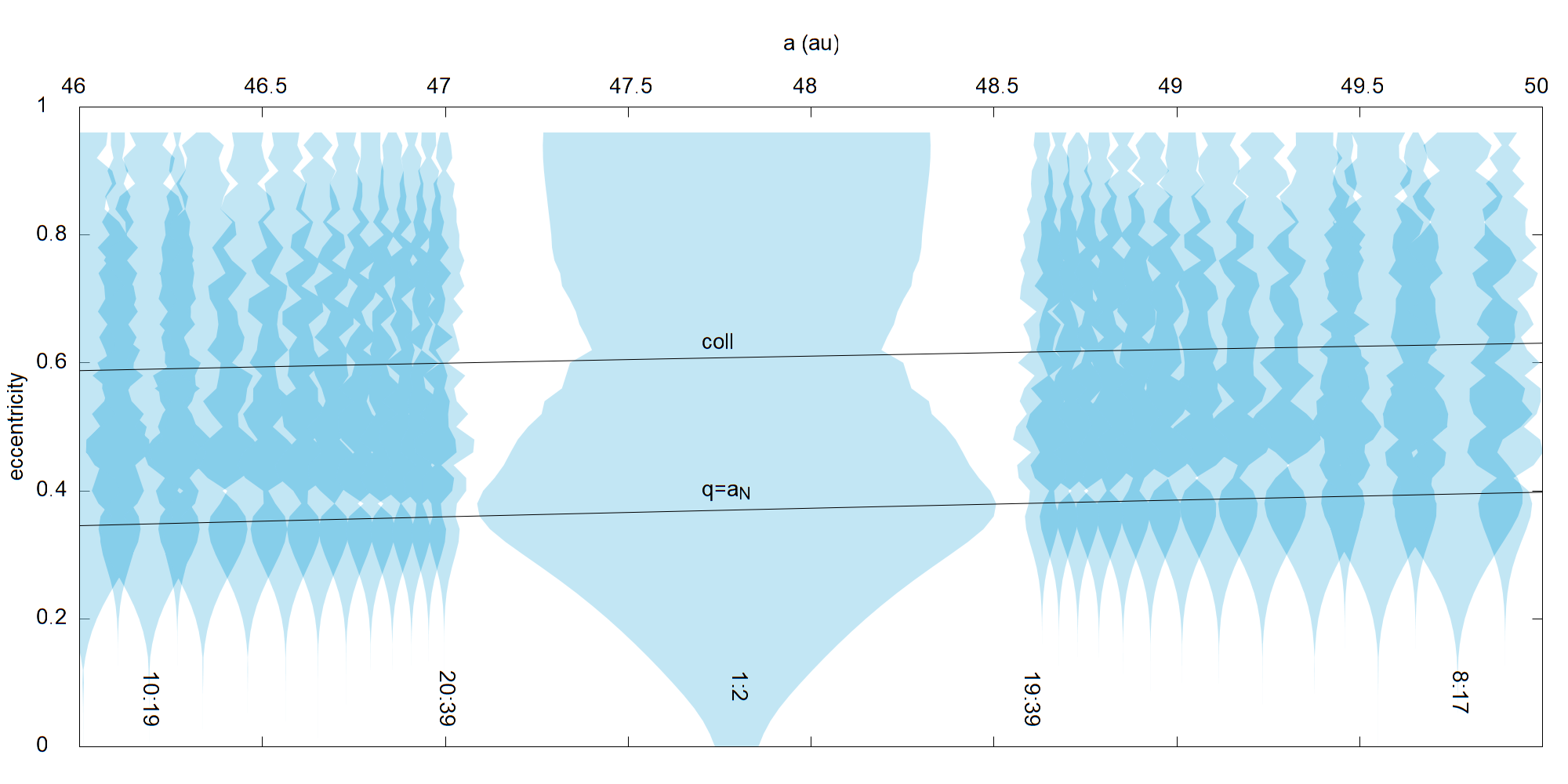}
	\caption{Detail of Fig.  \ref{atlasi10} showing the region close to resonance 1:2 considering all resonances with
		$k\leq 40$ and  $k_p\leq 40$. Darker regions correspond to superposition of resonances.}
	\label{atlasi10a4550}     
\end{figure}

One may guess that the isolation of  resonances 1:$k$ shown in the preceding figures is only apparent and due to the limits imposed to $k_p,k$. 
Probably, going to greater $k_p$ and $k$, resonances 1:$k$ will be surrounded with closer resonances as is the case of resonances 2:$k$ (see Fig.  \ref{atlasi10} for example). 
In order to study this point we calculated the resonances approaching the resonance 1:2 from the left and right sides up to $k_p,k=40$ and  Fig.  \ref{atlasi10a4550} shows the result. 
Note the sinusoidal-like variations of the widths conforming the eccentricity increases. They are related to successive  changes in the stability of the equilibrium points or to the alternation between the location of the principal and the secondary minimum of $\mathcal{R}$. A similar behavior is shown in figure 1 by \cite{Lan2019}.
 Fig.  \ref{atlasi10a4550} shows that  the resonance 1:2 is now more threatened by weak neighbor resonances but the doubt persists: is it in fact isolated?
Are these high $k_p$:$k$ resonances real?

\begin{figure}
	\includegraphics[width=1\textwidth]{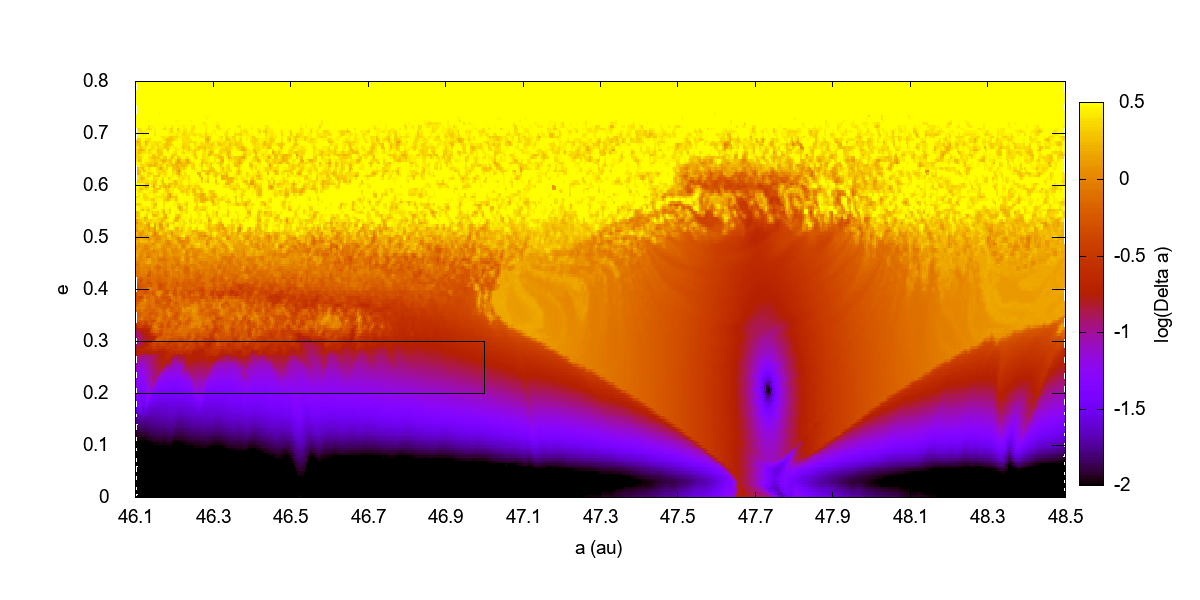}
	\caption{Dynamical map for the resonance 1:2 and its neighborhood  obtained integrating the Sun, the giant planets in coplanar orbits with initial $\lambda_N \sim 0^{\circ}$ and massless test particles during 200 orbital periods with $i=10^{\circ}$ and initial $\omega=90^{\circ}$, $\Omega=0^{\circ}$ and $M=95^{\circ}$ ($\sigma_{1:2}=280^{\circ}$).
		 Horizontal and vertical axis are  the initial baricentric semimajor axis and eccentricity respectively. The logarithmic color scale shows maximum detected variations in baricentric $a$ in au. Compare with Fig.  \ref{atlasi10a4550}. High $k_p$:$k$ weak resonances with Neptune are present at the left and the region defined by the rectangle is augmented in Fig.  \ref{frag}.  The resonance 1:4 with Uranus appears at the right side at $a\sim 48.4$ au.		 
		   }
	\label{efjsun10}     
\end{figure}

To solve the point we appeal to dynamical maps. We integrated the Sun with the giant planets in its present orbits but assuming $i=0^{\circ}$ for all of them and 80000 particles with initial $45.5<a<48.5$ au, $0<e<0.8$ and  all them with the same initial $\Omega=0^{\circ}$, $\omega=90^{\circ}$, $i=10^{\circ}$ and mean anomaly $M=45^{\circ}$. Each particle is integrated by 200 orbital periods and the detected changes in baricentric $a$ are plotted in logarithmic scale.
Fig.  \ref{efjsun10} shows the resulting map.
Large $\Delta a$ (in yellow) are due to close encounters or highly chaotic evolutions. Very small $\Delta a$ (in black and blue) are due to secular evolutions. Regular structures inside the secular or chaotic  regions are due to oscillations in $a$ due to resonances.
The domain of the resonance 1:2 is clear  and their limits are neatly defined.
The  yellow fuzzy horizontal band is generated by intersections and to close encounters with Neptune and Uranus. 
For particles with orbits coplanar with the planets, all the region above the line defined by $q=a_N$ (that is $e>0.36$) would be
 yellow.  Particles with eccentricities close to the collision bands defined by Uranus and Neptune are unstable because  their circulating $\omega$ eventually will take a value close to the one given by Eq. (\ref{col}) corresponding to a collision with one of the planets. The initial conditions were taken so that initial $\sigma \sim 280^{\circ}$ for the resonance 1:2 which guarantees large amplitude  librations and maximum resonance widths in the dynamical map. But for eccentricities close to the collision curve this initial condition produces close encounters with Neptune disrupting the resonance. That is why the resonance does not persist close to the collision curve. 

\begin{figure}
	\includegraphics[width=1\textwidth]{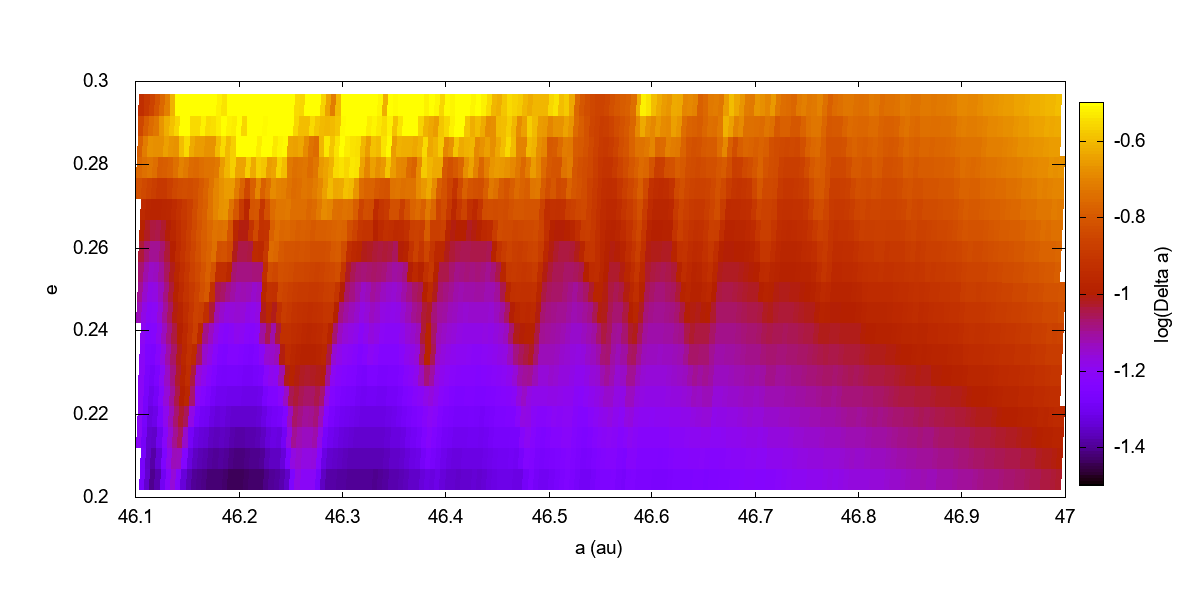}
	\caption{Zoom of the rectangle of Fig.  \ref{efjsun10} showing from left to right  with fading intensity the resonances 10:19 at $a\sim 46.13$ au, 11:21, 12:23, 13:25, 14:27 and 15:29 at $a\sim 46.65$ au.}
	\label{frag}     
\end{figure}

What we want to stress is that there are several resonances at the left of the resonance 1:2 but they do not affect the limits of the resonance. Below the yellow unstable band we can identify traces of the resonances up to $a\sim 46.7$ au corresponding to resonance 15:29. We show a zoom of this region at Fig.  \ref{frag}. Between this resonance and resonance 1:2 there is no dominant resonant structures and they do not affect the borders of resonance   1:2. Note that,  at the right of the resonance 1:2 in Fig.  \ref{efjsun10}, it is possibly to distinguish a ghostly pattern due to the resonance 1:4 with Uranus.  
Above the yellow band,
 the continuous chaotic region  is due to close encounters with the planets mostly, not to superposition of resonances. 
 We have also calculated dynamical maps extending the integration time  and no new resonances show up. We also obtained dynamical maps for polar orbits of $i=90^{\circ}$ that show a very rich resonant structure between the collision lines with Neptune and Uranus but no invading resonances appear close to resonance 1:2.

We can understand why this resonance is isolated noting that
 the nearest resonances are those with large $k_p,k$ which means that they are weak but fundamentally  that for the dynamical start-up of the resonances it is necessary a large number of orbital revolutions dropping the efficiency of the resonant mechanism. 
The resonant mechanism works because there is a sequence of  perturbations by Neptune that is repeated after $k_p$ revolutions of the particle. During that time interval both objects, planet and TNO, cannot change very much their orbits otherwise the sequence of perturbations will be broken and the resonance cannot be installed. The 
greater the number of perihelion passages, $k_p$, that the TNO must complete
 the greater the probability that its orbit change due to the cumulative effect of planetary perturbations. 
 We can guess that this situation of isolation of resonance 1:2 is  analogue for the other 1:$k$ resonances
 because their potentially threatening resonances are also those with large
  $k_p$ and $k$ values.
 Therefore, it is reasonable that there are some maximum values for $k_p,k$  so that the real resonances can work.
 In the specific case of the resonance 1:2, the closest resonance with some, though tiny, dynamical traces seems to be 
15:29. 
\cite{Yu2018} performed numerical integrations of particles with $30<a<100$ au looking for captures in MMRs and they obtained
a very illustrative map of captures for resonances $k_p$:$k$ ($q$:$p$ using their notation) that shows that the efficiency of captures drops substantially for $k_p>13$ in very good agreement with our results.
It is worth mentioning that \cite{Chambers1997} studied the credibility of exterior MMRs with Jupiter and concluded that for high $k/k_p$ ratios the binding energies of the comets to the Sun are not large enough to overcome the planetary perturbations and the resonances cannot work. Our case is different, it is not a problem of low binding energy (all orbits have $a\sim 47$ au in the case of resonance 1:2) but of the large number of orbital revolutions that the TNO needs to complete in order to set up the resonant mechanism.
Remember that we have also showed that for large $k_p$ the fragility of the resonance is large. Then,
although in this work we have not investigated the problem in depth, there are enough evidence indicating there is a maximum limit for $k_p$ (close to $\sim 14$)  in the TNR  so that a resonance can be installed.

\section{Six resonant populations of TNOs}
\label{respop}

Using the orbital elements obtained from AstDyS\footnote{newton.spacedys.com/astdys}  by june 2019 we performed numerical integrations using EVORB \citep{Fernandez2002} of the four giant planets plus the TNOs with semimajor axes close to the resonances 1:1, 2:3, 3:5, 4:7, 1:2 and 2:5. We automatically analyzed the output of the first $10^5$ years searching for librations of the corresponding critical angles. The automatic detection is based on a statistical analysis of the critical angle. If the distribution of the calculated $\sigma$ is approximately uniform between 0$^{\circ}$ and 360$^{\circ}$ we discard the possibility of being a resonant TNO. But, if there is an obvious  concentration in some interval and its semimajor axis is inside the limits of the resonance we consider the object as resonant. In cases that are not very conclusive we checked its status by direct inspection. We identified 652 TNOs evolving in these resonances.

\begin{figure}
	\includegraphics[width=1\textwidth]{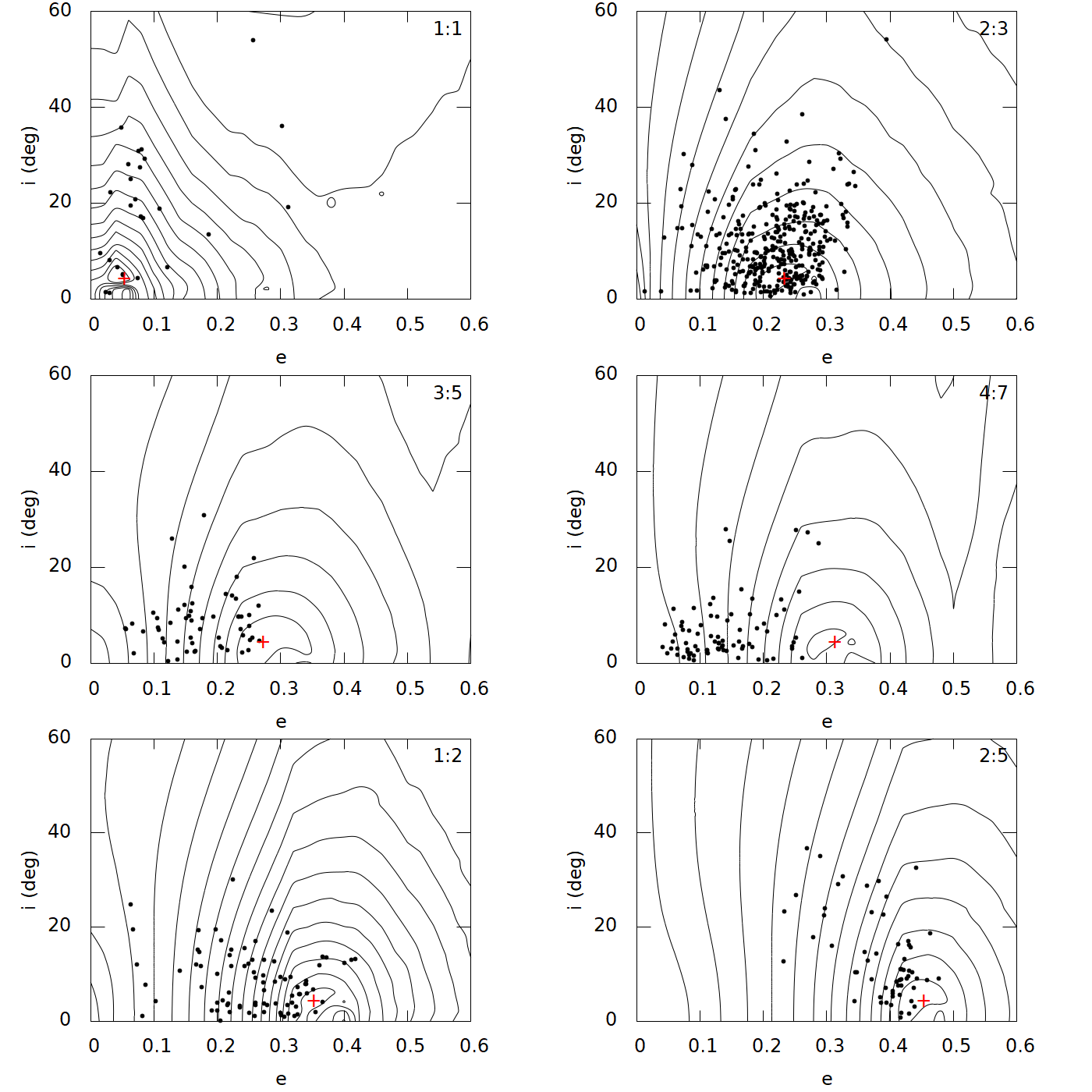}
	\caption{Six populations of resonant objects represented by black dots accompanied with level curves of maximum widths in au in steps of 0.1 au. The location corresponding to the maximum width in all the space considered is indicated by a red cross.}
	\label{6popu}     
\end{figure}

Fig.  \ref{6popu} shows the 6 populations of resonant objects inside level curves of maximum widths for each resonance. The level curves are the same that can be deduced from the corresponding Figs.  \ref{1to1f} to \ref{2to5f}.
These are the maximum widths corresponding to stable librations obtained when varying $\omega$ between 0$^{\circ}$ and 90$^{\circ}$.
Maximum widths in the plane $(e,i)$ for low inclination orbits corresponds to orbits with $q\sim a_N$, that means $e\sim e_c$. Then it is natural that the populations tend to concentrate close to the maximum width regions but below that eccentricities, avoiding collisions with Neptune.
All populations are nearly concentrated close to the region $(e,i)$ where the maximum width are located with the notable exception of population  4:7 and in a lesser extent also 3:5. These populations appear shifted to the left, to lower eccentricities in comparison with the other populations. 
\cite{Lykawka2005} studied the 4:7 resonance and found that the most stable region is approximately defined by $0.25<e<0.3$ in good agreement with the location of the region corresponding to  maximum widths we show in Fig.  \ref{6popu}. Nevertheless, the mean eccentricity for the 4:7 population is 0.14,
well below the most stable region. 

\begin{figure}
	\includegraphics[width=1\textwidth]{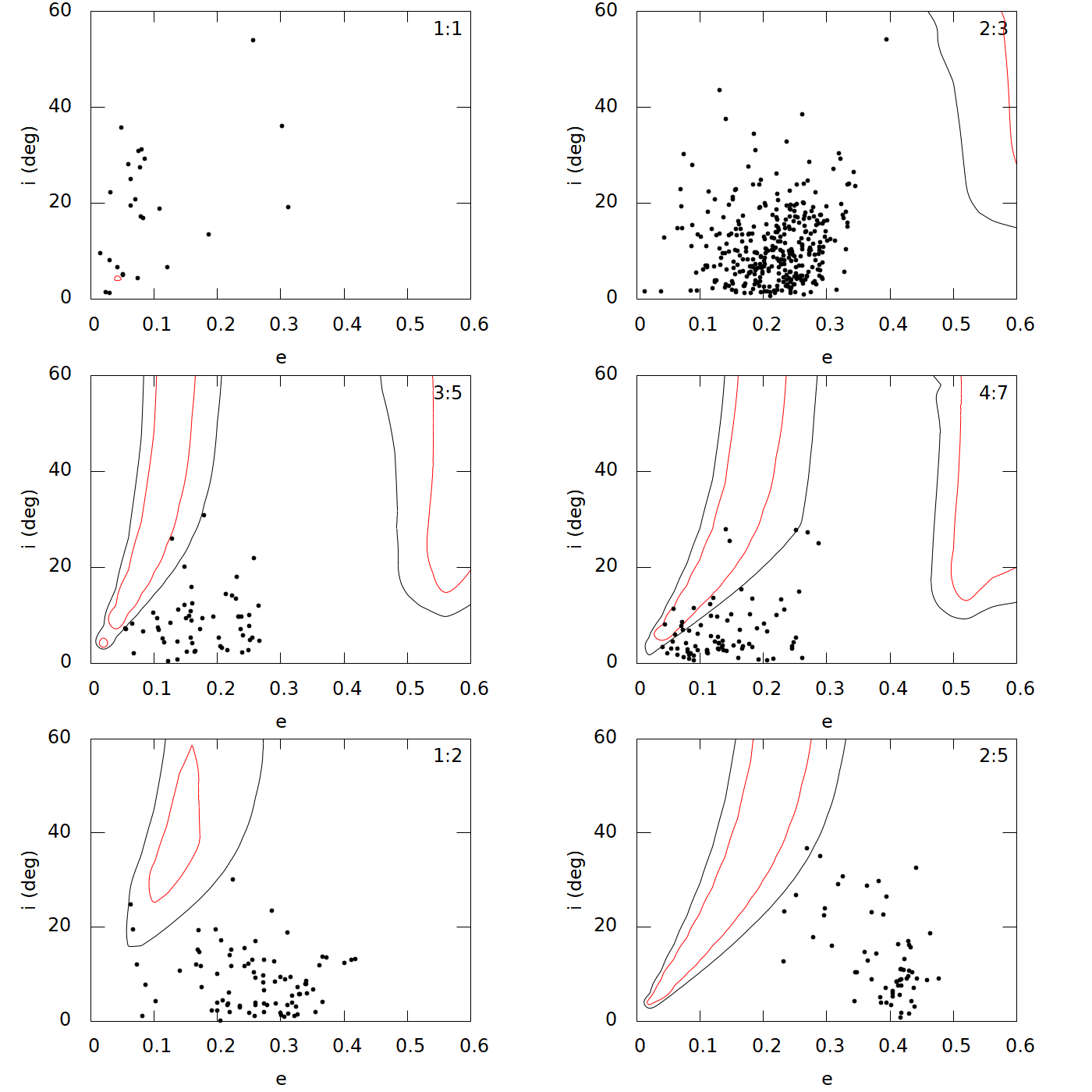}
	\caption{Same populations of Fig.  \ref{6popu} but showing the resonances' fragility. Black level curves indicate $f=0.5$ and red level curves $f=1$. All shown region of resonance 1:1 has $f<0.5$.}
	\label{6popufr}     
\end{figure}

Fig.  \ref{6popufr} shows the six population along with level curves of fragility.
In general each population is
  inside regions of $f<0.5$ and far from the region of $f=1$ represented by red lines with the exception, again, of the populations 3:5 and 4:7 which are shifted  to lower eccentricities, to regions of larger fragility.
 We have calculated an atlas of resonances near the resonance 4:7 and going up to values of $k_p,k\leq 40$ we  find resonances that invade both borders of the resonance. But, by means of dynamical maps we verified that these threatening resonances  in fact  do not exist because of their impossibility to be installed, as in the case of the resonance 1:2 that we have studied.  Then it is not clear for us whether the anomalous low eccentricities in the 3:5 and 4:7 resonant populations are generated by the resonances themselves, by secular effects inside the resonances or by cosmogonic reasons.

\section{Summary}
\label{discu}

We have developed a simple model for the calculation of librations centers, periods and widths of arbitrary MMRs with a planet in circular orbit with no restrictions about the orbital elements of the small body.  No series developments are necessary and we provide a code to calculate the resonance's properties.
For the calculation of the maximum widths we follow the idea of stable librations. For its computation we adopt the criteria of rejecting perturbations generated at planetocentric distances lower than $3R_{H}$. The obtained widths are in good agreement with results from numerical integrations and from other authors. 
We  have not investigated in deep the reasons why the resonances become unstable at the borders for some regions in the space $(e,i)$, but we have found several cases of low $k_p$:$k$ resonances in the TNR where the instability is caused by encounters with the planet, not due to the superposition of resonances.

We showed the relevance of $\omega$ in defining the properties of the resonances, like the location of the libration centers and the resonance widths.
Considering the time variation of $\omega$, we  introduce the concept of fragility of the resonances which is a measure of how much the resonance width can change while varying $\omega$ but preserving $(e,i)$. The fragility is irrelevant for zero inclination or zero eccentricity orbits, but in other cases is  important  and we showed that in the TNR for resonances with greater $k_p$ the corresponding fragilities are greater.
A  resonance with high fragility in some region of $(e,i)$ will not be able to sustain resonant TNOs for long time scales in that region.

The model allowed us to present a very complete atlas of resonances beyond Neptune that shows that resonances 1:$k$ and 2:$k$ are the strongest and most isolated ones even for polar and retrograde orbits, confirming the findings of other previous studies  \citep[for example]{2006Icar..184...29G,2007Icar..192..238L,Yu2018,2019Icar..317..121G,Lan2019}.
 Their isolation is related to the impossibility that the neighbor  high $k_p$:$k$ resonances can be installed. These high $k_p$:$k$ resonances, probably partially overlapping each other, only exist in theory. 
We also found that high inclination resonances become wider near the collision curve, a fact that  could facilitate the capture in resonance for high inclination objects.

We studied six resonant populations of TNOs and we found that in general these populations tends to concentrate in the region of the plane $(e,i)$ where the resonance widths are larger and the resonances less fragile. But there is a notable exception which is the population inside the resonance 4:7 which is clearly shifted to lower eccentricities. In a lesser extent the resonance 3:5 shows a similar behavior. We do not have an explanation for this particular behavior of these resonances.

All evidence points to resonances 1:$k$ and 2:$k$ as the strongest, widest, less fragile and more isolated resonances in the TNR for all interval of inclinations and eccentricities. The evidence also points to a limit vale of $k_p$, may be 14, so that a resonance can be installed in the TNR.

\begin{acknowledgements}
The author acknowledges the work of the reviewers. Support from PEDECIBA and SNI
are also acknowledged.
\end{acknowledgements}

 \section*{Conflict of interest}

 The authors declare that they have no conflict of interest.

\bibliographystyle{spbasic}      
\bibliography{tnotaba}   

\end{document}